\documentclass[11pt]{article}
\input epsf
\usepackage{amssymb, latexsym, amsmath, psfig}

\makeatletter \@addtoreset{figure}{section}
\def\thefigure{\thesection.\@arabic\c@figure}
\def\fps@figure{h, t}
\@addtoreset{table}{bsection}
\def\thetable{\thesection.\@arabic\c@table}
\def\fps@table{h, t}
\@addtoreset{equation}{section}

\newtheorem{theorem}{Theorem}[section]
\newtheorem{proposition}[theorem]{Proposition}
\newtheorem{lemma}[theorem]{Lemma}
\newtheorem{corollary}[theorem]{Corollary}
\newtheorem{remark}[theorem]{Remark}
\newtheorem{definition}[theorem]{Definition}
\newfont{\tenbi}{cmbxti10}

\def\la {\lambda}

\begin{document}
\title{ Algebraically closed real geodesics
on n-dimensional ellipsoids are dense in the parameter space
and related to hyperelliptic tangential coverings \footnote{AMS Subject Classification 37J35,
70H12,70H06}}

\author {Simonetta Abenda \\
Dipartimento di Matematica e CIRAM \\
Universit\`a degli Studi di Bologna, Italy \\
{\footnotesize  abenda@ciram.unibo.it}
 } \maketitle
\begin{abstract}
The closedness condition for real geodesics on $n$--dimensional
ellipsoids is in general transcendental in the parameters
(semiaxes of the ellipsoid and constants of motion). We show that
it is algebraic in the parameters if and only if both the real and
the imaginary geodesics are closed and we characterize such
double--periodicity condition via real hyperelliptic tangential
coverings. We prove the density of algebraically closed geodesics
on $n$--dimensional ellipsoids with respect to the natural
topology in the $2n$--dimensional real parameter space. In
particular, the approximating sequence of algebraic closed
geodesics on the approximated ellipsoids may be chosen so to share
the same values of the length and of the real period vector as the
limiting closed geodesic on the limiting ellipsoid.

Finally, for real doubly--periodic geodesics on triaxial
ellipsoids, we show how to evaluate algebraically the period
mapping and we present some explicit examples of families of
algebraically closed geodesics.
\end{abstract}

\section{Introduction}

Integrability of the geodesic motion on a triaxial ellipsoid $Q$
was proven in 1838 by Jacobi \cite{J} who reduced the system to
hyperelliptic quadratures; moreover Weierstrass \cite{Weier}
integrated the system in terms of theta--functions on a genus 2
hyperelliptic curve. The geodesic flow has many interesting
geometric properties: in particular, each geodesic on $Q$
oscillates between the two lines of intersection of $Q$ with a
confocal hyperboloid $Q_c$ (caustic) and by a theorem by Chasles
\cite{Chasles} all the tangent lines to the geodesics are also
tangent to $Q_c$. The generic geodesic is quasi--periodic and, in
case a geodesic on $Q$ is closed, then all the geodesics on $Q$
tangent to the same confocal hyperboloid are also closed and have
the same length.

The theorem of Chasles generalizes to $n$--dimensional quadrics
$Q$ and the set of common tangent lines to $n$ confocal quadrics
plays an important role in the study of the geodesics on any of
such quadrics and in the reformulation of integrability of the
system in the modern language of algebraically integrable systems
(see Moser \cite{Moser,Moser2}, Kn\"orrer \cite{Knorr,Knorr2} and
Audin \cite{Audin}). In particular, in \cite{Knorr2} Kn\"orrer
settled the so--called Moser--Trubowitz isomorphism between the
geodesics on quadrics and the stationary solutions to the Korteweg
de Vries equation (KdV).

One of the consequences of Chasles theorem is that, when a
geodesic on $Q$ is closed, all the geodesics sharing the same
values of the constants of motion are closed and of the same
length (see for instance \cite{Klin}). The condition for a
geodesic on an $n$--dimensional quadric $Q$ to be closed is then
expressed as a certain linear combination of integrals of
holomorphic differentials on a hyperelliptic curve. Such condition
is transcendental in the parameters of the problem (semiaxes of
the quadric $Q$ and parameters of the caustics) and, by the
Moser--Trubowitz isomorphism, it is equivalent to impose that the
stationary solutions of the KdV are real periodic in $x$.

\medskip

\paragraph{Characterization of the set for which the closedness property
of real geodesics on $n$--dimensional ellipsoids is algebraic in
the parameters} A natural question is then: is it possible to
settle extra conditions so that the closedness property
(\ref{realp}) of the geodesic be algebraic in the parameters
(semiaxes of the quadric $Q$ and the constants of motions)? In
\cite{Fed05, AF}, we found a set of sufficient conditions in the
complex setting: we introduced and characterized a family of
algebraic closed geodesics associated to hyperelliptic tangential
covers.

\medskip

The results in the above papers indicate that the closedness
property is algebraic (in the parameters) if the periodicity
condition is essentially one--dimensional in the complex setting.
In the algebraic-geometric setting, this in turn means that the
closedness condition is algebraic (in the parameters) if it is
equivalent to the inversion of an elliptic integral.

\medskip

In the present paper, we complete the characterization of
algebraically closed geodesics: we restrict ourselves to the real
setting and we settle the necessary and sufficient conditions so
that the closedness property be algebraic in the real parameters
(semiaxes of the ellipsoid and constants of motion).

In particular, we prove that the closedness condition is algebraic
in the parameters if and only if both the real and the imaginary
geodesics on the $n$--dimensional ellipsoid are closed. The double
periodicity condition we introduce here for the real geodesics on
ellipsoids is modelled after a similar condition  by Mc Kean and
van Moerbeke\cite{MKVM} for the real Hill problem.

Then we explicitly show that the double--periodicity condition is
equivalent to the existence of a real hyperelliptic tangential
cover \cite{Tr,TV,TV2}, thus completing the study started in
\cite{Fed05,AF}.

The conclusion is then the following: the closedness property is
algebraic in the parameters of the problem (square semiaxes of the
ellipsoid and constants of the motion) if and only if the
double--periodicity condition holds and, in such a case, the
closedness property is equivalently expressed by an elliptic
integral associated to the elliptic curve in the hyperelliptic
tangential covering.

\medskip

We remark that the appearance of hyperelliptic tangential covers
is natural, since their role in the topological classification of
elliptic KdV solitons in the complex moduli space of hyperelliptic
curves is well known after Treibich-Verdier\cite{TV}-\cite{Ver}
and the Moser-Trubowitz isomorphism ensures a relation with the
geodesic problem.

Since the classification of closed geodesics on real quadrics (and
of real KdV elliptic solitons) are of a certain interdisciplinary
interest and the double-periodicity property of geodesics on
ellipsoids is not invariant under general birational
transformations, we explicitly describe such coverings for the
geodesic problem. In particular, we investigate the real structure
of the elliptic curve of the covering and we show that the
associated lattice is rectangular ({\it i.e.} all of the
ramification points of the elliptic curve are real).

\medskip

We remark also that it is appropriate to call doubly-periodic the
geodesics associated to hyperelliptic tangential covers, since the
coordinates and momenta are doubly--periodic in the length
parameter $s$, that is they are expressed in terms of elliptic
functions of $s$; moreover it is also appropriate to call
algebraic the doubly-periodic geodesics, since the closedness
property is algebraic in the parameters (semiaxes of the ellipsoid
and constants of motion).

\medskip

\paragraph{The density property}
The second set of questions we characterize in the present paper
concerns the density characterization of algebraically closed
geodesics. We show that it is possible to approximate a given real
closed geodesics on a given ellipsoid with a sequence of real
algebraically closed geodesics on perturbed ellipsoids with
perturbed constants of motion. Moreover, such approximate
algebraically closed geodesics may be chosen so to share the same
length and/or the same value of the period vector as the limiting
geodesic.

Our estimates are optimal in the sense that we are able to count
the number of parameters which may be kept fixed in the
approximation process of real closed geodesics on a given
$n$--dimensional ellipsoid via a sequence of doubly--periodic real
closed geodesics on perturbed ellipsoids.

For instance, in the simplest case (geodesics on triaxial
ellipsoids), Theorem \ref{optimal1} implies that we may keep fixed
one parameter: indeed there are four parameters (the three
semiaxes and the caustic parameter), two conditions (length and
period mapping of the real closed geodesic to be approximated
algebraically) and one extra condition (the approximating
geodesics satisfy the double--periodicity condition, {\it i.e.}
the period mapping of the associated imaginary geodesic has to be
rational). Similarly Theorem \ref{optimal2} implies that we may
keep fixed two parameters if we allow the length of approximating
algebraic geodesic to vary a little.

The density characterization follows from a theorem by McKean and
van Moerbeke \cite{MKVM} which allows the construction of a
locally invertible analytic map from the set of the parameters of
the problem (the semiaxes and the caustic parameters) to the
quasi-periods associated to the geodesics on $n$--dimensional
ellipsoids.

\medskip

\paragraph{The case of triaxial ellipsoids: the period mapping and the examples}
We then specialize to the case $n=2$ (triaxial ellipsoids), where
a more detailed characterization of doubly periodic geodesics is
possible since the associated two dimensional (complex) torus is
isogenous to the product of two elliptic curves. The first
elliptic curve is associated to the hyperelliptic tangential
covering, while the properties of the second covering have been
discussed by Colombo {\it et al.} \cite{CPP}.

In particular, we show that the period mapping of a doubly
periodic real geodesic is algebraic in the parameters of the
problem and may be computed using the topological character of the
second covering.

We also work out the reality condition for geodesics on triaxial
ellipsoids associated to degree $d=3,4$ hyperelliptic tangential
covering and we compute the period mapping using the topological
character of the associated second covering.

Finally, we prove the existence of real doubly-periodic geodesics
associated to the one--parameter family of degree two coverings
with the extra automorphism group $D_8$ \cite{Igusa,AP}. In this
case the two elliptic curves of the covering are isomorphic, i.e.
they have the same $j$--invariant, and the geodesics are
doubly-periodic for a dense set in the parameter space. In view of
the above discussion, for such values of the parameter, the given
hyperelliptic curve also admits another cover which is
hyperelliptic tangential of degree $d>2$ (see Figure 4 for an
explicit example).

\medskip

\paragraph{The plan of the paper}
The plan of the paper is the following: in the next section we
summarize some well known facts about geodesics on
$n$--dimensional real quadrics. In section 3, we introduce
doubly--periodic closed geodesics, hyperelliptic tangential covers
and we characterize the algebraic condition of closedness; in
section 4 we present the density results; in sections 5 and 6 we
specialize to algebraic closed geodesics on triaxial ellipsoids,
we characterize algebraically the period mapping and we present
the examples.

Since the classification of closed geodesics on real quadrics and
of real KdV elliptic solitons are of a certain interdisciplinary
interest, we have tried at best to report our results in a way
comprehensible also to not experts in the theory of Riemann
surfaces.

\section{Closed geodesics on ellipsoids}
The Jacobi problem of the geodesic motion on an $n$-dimensional
ellipsoid
\[
Q \; : \; \left\{ \frac{X_1^2}{a_1} +\dots +
\frac{X_{n+1}^2}{a_{n+1}} =1 \right\}
\]
is well known to be integrable and to be linearized on a covering
of the Jacobian of a genus $n$ hyperelliptic curve (see
\cite{Moser}). Namely, let $l$ be the natural parameter of the
geodesic and $\la_1,\dots,\la_n$ be the ellipsoidal coordinates on
$Q$ defined by the formulas
\begin{equation}
\label{spheroconic2} X_i=\sqrt{ \frac{(a_i-{\la}_1)\cdots
(a_i-{\la}_n)} { \prod_{j\ne i}(a_i-a_j)} }, \qquad i=1,\dots,
n+1.
\end{equation}
Then, denoting $V_i = \dot X_i = dX_i / dl$, $i=1,\dots,{n+1}$ and
$\dot\lambda_k= d\lambda_k/d l$, $k=1,\dots, n$, the corresponding
velocities, the total energy $\displaystyle \frac{ 1}{2}
(V_1^2+\cdots+V_{n+1}^2)$ takes the St\"ackel form
\begin{gather*}
H = -\frac{1}{8}\sum^{n}_{k=1} \lambda_k{\dot\lambda}^{2}_{k} \,
\frac{ \prod\limits^{n}_{j\ne k} (\lambda_{k}-\lambda_{j})
}{\prod\limits_{i=1}^{n+1} (\lambda_{k} -a_i)} .
\end{gather*}
According to the St\"ackel theorem, the system is Liouville
integrable. Upon fixing the constants of motion
$H=h_1,c_1,\dots,c_{n-1}$ and after the re-parametrization
\begin{equation} \label{tau-1}
dl =\la_1\cdots\la_n\, \frac{ds}{\sqrt{8h_1}} ,
\end{equation}
the evolution of the $\lambda_{k}$ is described by quadratures
which involve $n$ independent holomorphic differentials on a genus
$n$ hyperelliptic curve\footnote{for the necessary definitions and
classical properties of hyperelliptic curves we refer to
\cite{Far, Gunn}} whose affine part takes the form
\begin{equation}\label{gamma}
\Gamma \; : \;\; \big\{\mu^2= -\lambda \prod_{i=1}^{n+1} (\lambda
- a_i) \prod_{k=1}^{n-1} (\lambda-c_k) \big\} = \big\{\mu^2=
-\prod_{i=0}^{2n} (\lambda - b_i) \big\} ,
\end{equation}
where we set the following notation throughout the paper
\begin{equation}\label{param}
\{ \, 0, a_1<\dots< a_{n+1}, c_1<\cdots< c_{n-1} \, \} = \{ b_0=0<
b_1< \cdots< b_{2n} \}.
\end{equation}

\begin{remark}\label{rem1} {\rm
Following \cite{Knorr,Audin}, the reality condition for geodesics
on ellipsoids is equivalent to either $c_i = b_{2i}$ or $c_i =
b_{2i+1}$, $i=1,\dots,n-1$.

A first consequence is that, given the ellipsoid $Q$ the values of
the real constants of motion $c_i$s can't take arbitrary values.
As an example, in the simplest case $n=2$ (triaxial ellipsoid),
given the square semiaxes $0<a_1<a_2<a_3$ the real constant of
motion $c$ satisfies either $a_1<c<a_2$ or $a_2<c<a_3$.

On the other side, it also implies that given a $(2n)$-tuple
$0<b_1< \cdots< b_{2n}$ ({\it i.e.} given the hyperelliptic curve
$\Gamma$), there are a finite number of mechanical configurations
associated to it. For instance, again in the simplest case $n=2$,
to the $4$-tuple $b_1<b_2<b_3<b_4$ there are associated either the
geodesics with constant of motion $c=b_2$ on the ellipsoid with
square semiaxes $b_1<b_3<b_4$ or the geodesics with constant of
motion $c=b_3$ on the ellipsoid with square semiaxes
$b_1<b_2<b_4$.}
\end{remark}

\begin{remark}\label{rem2} {\rm Throughout the paper, for any
given curve $\Gamma$ with all real branch points as in
(\ref{gamma}), we use the following basis of holomorphic
differentials
\begin{equation}\label{basis}
\omega_j =\frac{\la^{j-1} d\la }{w}, \quad\quad j=1,\dots, n,
\end{equation}
and the homological basis $\alpha_i,\beta_i$, $i=1,\dots,n$ (see
Figure 1), so that the periods $\oint_{\alpha_i}\omega_j \in
{\mathbb R}$, $i,j=1,\dots,n$.}
\end{remark}

Then, the quadrature gives rise to the Abel--Jacobi map of the
$n$-th symmetric product $\Gamma^{(n)}$ to the Jacobian variety of
$\Gamma$,
\begin{equation}\label{AB}
\int \limits^{P_1}_{P_0} \omega_j + \cdots + \int
\limits^{P_n}_{P_0} \omega_j =\Bigg \{
\begin{aligned}
s + const.,\quad \mbox{ for } & j=1\, , \\
const., \quad \mbox{ for } & j=2, \dots, n ,
\end{aligned} \end{equation}
where and $P_0$ is a fixed basepoint and $ P_k=\left(\lambda_k,w_k
\right) \in \Gamma$, $k=1,\dots, n$. Then, the geodesic motion in
the new parametrization is linearized on the Jacobian variety of
$\Gamma$. Its complete theta-functional solution was presented in
\cite{Weier} for the case $n=2$, and in \cite{Knorr} for arbitrary
dimensions, whereas a topological classification of real geodesics
on quadrics was made in \cite{Audin}. In particular, the constants
of motion $c_1,\dots, c_{n-1}$ have the following geometrical
meaning (see \cite{Chasles,Moser}): the corresponding geodesics
are tangent to the quadrics $Q_{c_1},\dots , Q_{c_{n-1}}$ of the
confocal family $Q_c = \left\{ X_1^2/(a_1-c) +\dots +
X_{n+1}^2/(a_{n+1}-c) =1 \right\}$.

\begin{figure}[htb]
\centerline{\psfig{file=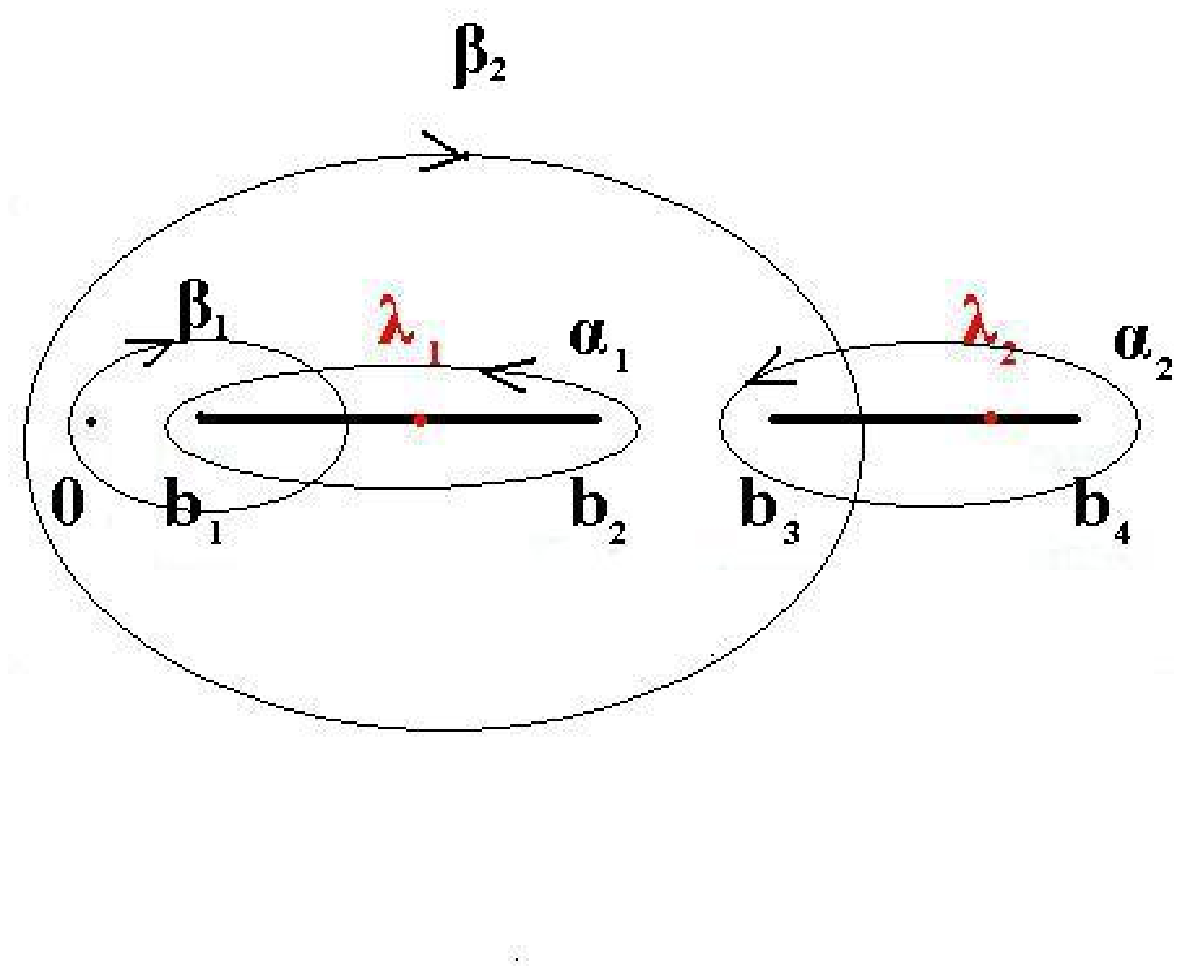,height=9cm,width=9.5cm}}
\vspace{-3.8 truecm} \centerline{Figure 1.}
\end{figure}

\paragraph{Closed geodesics and real Hill curves}
Let $\alpha_i,\beta_i$, $i=1,\dots, n$ be the conventional
homological basis depicted in Figure 1. Since we are interested in
the reality problem, it is not restrictive to take
$b_{2i-1}<\lambda_i<b_{2i}$, $i=1,\dots, n$, in the quadratures
(\ref{AB}). Then the real geodesic associated to (\ref{AB}) is
closed if and only if there exist non trivial $m_i\in {\mathbb
Z}$, $i=1,\dots,n$ and a real non vanishing $T>0$ such that
\begin{equation}\label{realp}
\displaystyle\sum_{i=1}^{n} m_i \oint_{\alpha_i} \omega_1
=T,\quad\quad \sum_{i=1}^{n} m_i \oint_{\alpha_i} \omega_j =0,
\quad j=2,\dots, n,
\end{equation}
where the basis of differentials has been introduced in
(\ref{basis}). From (\ref{realp}), it is self-evident that, if a
geodesic on $Q$ is closed, then all the geodesics sharing the same
constants of motion $c_1,\dots,c_{n-1}$ are closed. In the
following, we call Hill a hyperelliptic curve as in (\ref{gamma})
for which (\ref{realp}) holds.

As it is well known, Hill curves originally arose from the study
of isospectral classes connected with the periodic Korteweg--de
Vries equation (see \cite{Dub_Nov,IM,L,MKVM2,Nov,Krich,MKVM,
BEBIM}). Let ${\cal H}_{n}^{\mathbb R}$ be the real component of
the moduli space of the non singular genus $n$ hyperelliptic
curves with maximal number $n+1$ of connected components, so that
all the branch points are real and distinct,
$b_0=0<b_1<\dots<b_{2n}$, then (\ref{realp}) is equivalent to
require that $\Gamma$ is a real Hill curve up to the
Moser--Trubowitz isomorphism, which is associated to the
birational transformation $z=1/\lambda$ (which exchanges the
branch points at $0$ and $\infty$).

In particular, in \cite{MKVM}, it is proven that real Hill curves
are dense in the moduli space of curves ${\cal H}_{n}^{\mathbb
R}$. A similar statement holds true also for real closed
geodesics; however, since the set of equations (\ref{realp}) are
transcendental in the branch points of $\Gamma$, they are of
little use for the search of parameters corresponding to closed
geodesics. For real geodesics on ellipsoids, the above discussion
may be summarized in the following classical result

\begin{proposition}\label{1}
For any fixed choice of the square semiaxes $a_1< \dots < a_{n+1}$
and for any $n$-tuple $\zeta_i$, $i=1,\dots,n$ such that
$a_1\le\zeta_1 <a_2 <\cdots <a_n \le \zeta_n <a_{n+1}$, there is a
dense set $I\subset [\zeta_1,\zeta_2]\times \cdots \times
[\zeta_{n-1},\zeta_n]$ (in the natural topology of ${\mathbb
R}^{n-1}$), such that $\forall (c_1,\dots,c_{n-1})\in I$, there
exist nontrivial integers $m_i$, $i=1,\dots, n$ and a real $T>0$
such that (\ref{realp}) holds.
\end{proposition}

The above statement takes into account of the reality condition
settled in Remark \ref{rem1} and exhausts all possibilities
accordingly for real closed geodesics on $n$-dimensional
ellipsoids $Q$.

\section{Doubly--periodic closed geodesics, hyperelliptic
tangential covers and algebraic condition of periodicity in the
real parameter space} In the following we consider real geodesics
in the regular case when all square semiaxes and constants of
motion take distinct values. The periodicity condition
(\ref{realp}) is transcendental in the parameters of the problem
(the square semiaxes $a_1,\dots,a_n$ of the ellipsoid and
constants of motion $c_1,\dots,c_{n-1}$). So a natural question
is: is it possible to settle extra conditions so that the
periodicity condition (\ref{realp}) becomes algebraic in the
parameters?

In \cite{Fed05, AF}, we introduced and characterized a family of
algebraic closed geodesics associated to hyperelliptic tangential
covers in the complex setting. Fedorov \cite{Fed05} proved that
such geodesics are a connected component of the intersection of
the quadric $Q$ with an algebraic surface in $\mathbb R^n$. For
triaxial ellipsoids this surface is an elliptic or rational curve
and the explicit description of the algebraic surface in terms of
elliptic ${\cal P}$--Weierstrass functions in special cases of
such coverings was given in \cite{Fed05}. In \cite{AF}, we
computed the explicit expression of the coordinates $X_i(s)$ in
terms of one-dimensional theta-functions and applied such results
also to describe periodic orbits of an integrable billiard.

\smallskip

In this section, we complete the characterization of algebraically
closed geodesics, we restrict ourselves to the real setting and we
settle the necessary and sufficient conditions so that the
closedness property be algebraic in the real parameters (semiaxes
of the ellipsoid and constants of motion).

The conclusion is the following one: the periodicity condition
(\ref{realp}) is algebraic in the parameters of the problem if and
only if it is equivalent to the inversion of a single integral; by
Jacobi inversion problem the latter integral has to be elliptic.
The form of the periodicity condition implies that the elliptic
curve is the one associated to the hyperelliptic curve via the
hyperelliptic tangential covering. Finally, under our hypotheses,
${\cal E}$ has a real structure and we prove that the associated
lattice is rectangular ({\i.e.} all of the finite branch points of
${\cal E}$ are real).

% (since ${\rm Jac} (\Gamma)$ is isogenous to ${\cal E}\times
% {\cal A}_{n-1}$, with ${\cal A}_{n-1}$ an $(n-1)$--dimensional
% Abelian variety, there is a certain freedom in the construction of
% the covering).

\smallskip

Indeed, we introduce and characterize a double periodicity
condition for the geodesics on ellipsoids which is modelled after
a similar condition for the real Hill problem by Mc Kean and van
Moerbeke\cite{MKVM}. Then we explicitly show that this condition
is equivalent to the existence of a hyperelliptic tangential cover
(explicitly described in Definition \ref{setting2}).

The theorems \ref{fed} \cite{Fed05} and \ref{main} imply that the
periodicity condition (\ref{realp}) is algebraic in the parameters
of the problem if and only if the real closed geodesics are
doubly-periodic.

We remark that it is appropriate to call such geodesics
doubly-periodic, since the coordinates and momenta, $X_i(s),
V_i(s)$, $i=1,\dots, n+1$, are doubly--periodic in $s$, that is
they are expressed in terms of elliptic functions of $s$; moreover
it is also appropriate to call algebraic the doubly-periodic
geodesics, since the closedness property is algebraic in the
parameters (semiaxes of the ellipsoid and constants of motion).

\medskip

The appearance of hyperelliptic tangential covers is natural,
since their role in the topological classification of elliptic KdV
solitons in the complex moduli space of hyperelliptic curves is
well known after Treibich-Verdier\cite{TV}-\cite{Ver} and the
Moser-Trubowitz isomorphism ensures a relation with the geodesic
problem.

Since the double-periodicity property of geodesics on ellipsoids
is not invariant under general birational transformations (see
Lemma \ref{lemmabirat}), we explicitly describe such coverings for
the geodesic problem and we characterize their real structure.

\medskip

The plan of the section is the following: we first introduce the
double-periodicity condition and characterize it via a dual curve;
then we explicitly construct the hyperelliptic tangential cover
associated to the double--periodicity condition and give the
necessary and sufficient conditions so that the closedness
condition is algebraic in the parameters.

\medskip

\begin{definition} {\bf (The double periodicity condition) }
A hyperelliptic curve $\Gamma$ with real branch points as in
(\ref{gamma}) is associated to doubly--periodic closed geodesics
if and only if the real periodicity condition holds, that is there
exists a non trivial real cycle $\alpha = \sum_{i=1}^n m_i
\alpha_i$, such that
\begin{equation}\label{realp1}
\oint_{\alpha} \omega_1 =2 T,\quad\quad  \oint_{\alpha} \omega_j
=0, \quad j=2,\dots, n, \end{equation} and there exists a non
trivial imaginary cycle $\beta = \sum_{i=1}^n m^{\prime}_i
\beta_i$, such that
\begin{equation}\label{imp}
\oint_{\beta} \omega_1 =2 \sqrt{-1}T^{\prime},\quad\quad
\oint_{\beta} \omega_j =0, \quad j=2,\dots, n,
\end{equation}
for some non--zero real $T,T^{\prime}$.
\end{definition}

The conditions (\ref{realp1}) and (\ref{imp}) mean that both the
real and the imaginary geodesics on the ellipsoid $Q$ are closed.

\vskip .2 truecm

\paragraph{The dual curve}

To any given real curve $\Gamma$ like in (\ref{gamma}) McKean and
van Moerbeke \cite{MKVM} associate a dual real curve
$\Gamma^{\prime}$ with reflected branch points so that the real
KdV elliptic soliton is doubly--periodic in $x$ if and only if
both $\Gamma$ and $\Gamma^{\prime}$ are real Hill curves (for the
Hill operator). From the algebraic-geometric point of view, the
birational transformation which sends branch points of $\Gamma$
into those of $\Gamma^{\prime}$ is uniquely defined by the
requirement that it exchanges real and imaginary periods and it
transforms holomorphic differentials vanishing at the infinity
ramification point of $\Gamma$ to holomorphic differentials
vanishing at the infinity ramification point of $\Gamma^{\prime}$.
From the analytical point of view, the real solutions to the Hill
problem associated to the dual curve $\Gamma^{\prime}$ correspond
precisely to the imaginary solutions for the corresponding problem
on $\Gamma$.

We remark the nontriviality of this construction: birationally
equivalent curves are identified in the moduli space of
hyperelliptic curves; however, the topological characterization of
the real solutions to the Hill problem for KdV is not invariant
under general birational transformations. The same remark holds in
the case of closed geodesics.

\medskip

Below we apply the same idea to the case of the geodesic problem
on ellipsoids: the analogous construction maps the imaginary
geodesics on the ellipsoid $Q$ with constants of motion
$c_1,\dots,c_{n-1}$, to the real geodesics on a dual ellipsoid
$Q^{\prime}$ with constants of motion
$c_1^{\prime},\dots,c_{n-1}^{\prime}$. So the real geodesics on
$Q$ are closed and doubly periodic, if and only if both the real
geodesics on $Q$ and $Q^{\prime}$ are closed for the given
constants of motion.

To identify the dual curve $\Gamma^{\prime}$ ({\it i.e.} the dual
ellipsoid $Q^{\prime}$ and the dual constants of motion
$c_1^{\prime},\dots,c_{n-1}^{\prime}$), we recall that the
Moser-Trubowitz isomorphism exchanges the infinity ramification
point of the hyperelliptic curve associated to the classical Hill
problem, with the $(0,0)$ finite ramification point of the
hyperelliptic curve associated to the geodesic problem.

Moreover, either the real (\ref{realp1}) or the imaginary
(\ref{imp}) periodicity conditions for the geodesic problem are
equivalent to require that, given a hyperelliptic curve as in
(\ref{gamma}), there exists a nontrivial cycle $\gamma$ such that
$\oint_{\gamma} \omega =0$, for all holomorphic differentials
$\omega$ vanishing at the branch point $(0,0)$.

\begin{lemma}\label{lemmavan}
Let $\Gamma\, : \{ \mu^2 = -\prod\limits_{k=0}^{2n}
(\lambda-b_k)\}$ be as in (\ref{gamma}), let $P_0=(0,0)$ and let
$\omega_k= \lambda^{k-1}/\mu$, $k=1,\dots,n$, be the basis of
holomorphic differentials introduced in (\ref{basis}). Then the
holomorphic differential $\omega$ vanishes at the branch point
$P_0$ if and only if it is a linear combination of the holomorphic
differentials $\omega_2,\dots,\omega_n$.
\end{lemma}

{\it Sketch of the proof:} $\;\;$ Let $\tau$ be the local
coordinate in a neighborhood of $P_0=(0,0)$ such that
$\tau(P_0)=0$, then $\omega_k \approx A\tau^{2k-2}d\tau$,
$k=1,\dots,n$, where
$A=2\left(\sqrt{-\prod_{j=1}^{2n}b_j}\right)^{-1}
\quad\quad\square$.

\medskip

Finally, to construct the dual curve ${\Gamma}^{\prime}$ for the
geodesic problem, we must identify the birational transformations
which preserve the form of (\ref{gamma}) and transform holomorphic
differentials vanishing at the ramification point $(0,0)\in
\Gamma$ to holomorphic differentials vanishing at the ramification
point $(0,0)\in \Gamma^{\prime}$.

\begin{lemma}\label{lemmabirat}
The class of birational transformations between $\Gamma \, : \{
\mu^2 = -\prod\limits_{k=0}^{2n} (\lambda-b_k)\}$ and
$\Gamma^{\prime} \, : \{ \nu^2 = -\prod\limits_{k=0}^{2n}
(\rho-b_k^{\prime})\}$ which transform holomorphic differentials
vanishing at the ramification point $(0,0)\in \Gamma$ to
holomorphic differentials vanishing at the ramification point
$(0,0)\in \Gamma^{\prime}$ has the following two generators:
$\rho=\kappa \lambda$ and $\rho = b_1\lambda/(\lambda-b_1)$.
\end{lemma}

The first transformation is a homogeneous rescaling of all of the
parameters of the problem (square semiaxes and constants of the
motion) and it preserves the real periodicity condition. The
latter transformation is the analog of the one introduced by
McKean and van Moerbeke \cite{MKVM} for the periodic KdV problem
and it exchanges the real and the imaginary cycles.

Finally, the statement below gives a simple characterization of
doubly--periodic geodesics and is the analog of a theorem in
\cite{MKVM} for the Hill problem.

\begin{theorem}\label{theodual}
Let $\Gamma = \{\mu^2= - \lambda \prod\limits_{k=1}^{2n} (\lambda
-b_k)\equiv-\lambda \prod\limits_{i=1}^{n+1} (\lambda - a_i)
\prod\limits_{k=1}^{n-1} (\lambda-c_k)  \}$  be a real
hyperelliptic curve as in (\ref{gamma}) and $\Gamma^{\prime}\; :
\, \{ \nu^2 = -\prod\limits_{i=0}^{2n} (\rho-b_i^{\prime})
\equiv-\lambda \prod\limits_{i=1}^{n+1} (\lambda - a_i^{\prime})
\prod\limits_{k=1}^{n-1} (\lambda-c_k^{\prime})\}$ be the real
hyperelliptic curve whose branch points $b_i^{\prime}$ are related
to the $b_j$s by the birational transformation $\rho =
\displaystyle \frac{b_1\lambda}{\lambda-b_1}$.

Let $Q= \{ X_1^2/a_1+\cdots + X^2_{n+1}/a_{n+1} =1\}$ and
$Q^{\prime}= \{ X_1^2/a_1^{\prime}+\cdots +
X^2_{n+1}/a_{n+1}^{\prime} =1\}$.

Then the real geodesics associated on $Q$ with constants of motion
$c_1,\dots,c_{n-1}$ are doubly--periodic if and only if the real
geodesics respectively associated to $Q$ (with constants of motion
$c_1,\dots,c_{n-1}$) and to $Q^{\prime}$ (with constants of motion
$c_1^{\prime},\dots,c_{n-1}^{\prime}$) are closed.
\end{theorem}

In view of Remark \ref{rem1}, to any given $(2n)$-tuple
$b_1^{\prime},\dots,b_{2n}^{\prime}$, there are associated a
finite number of dual ellipsoids and dual constants of motion.
Clearly the Theorem \ref{theodual} implies that the real geodesics
on $Q^{\prime}$ be closed, for any admissible dual ellipsoid
$Q^{\prime}$ and constants of motion
$c_1^{\prime},\dots,c_{n-1}^{\prime}$ associated to
$b_1^{\prime},\dots,b_{2n}^{\prime}$.

In the last section, we apply Theorem \ref{theodual} both to
compute the period mapping associated to families of coverings and
to compute the parameters of Example \ref{lastexample}.

\medskip

\paragraph{Hyperelliptic tangential covers and the algebraic periodicity condition for
closed geodesics} In this paragraph, we prove that the
double--periodicity condition is necessary and sufficient for the
algebraic characterization of the closedness property of real
geodesics on $n$--dimensional ellipsoids. The statement follows
from the fact that the double periodicity condition settled by
equations (\ref{realp1}) and (\ref{imp}) is equivalent to the
existence of a real rectangular hyperelliptic tangential cover
defined in Definition \ref{setting2}.

\smallskip

Hyperelliptic tangential covers \cite{TV}-\cite{Tr} have
originally appeared in connection with the topological
classification of the $x$ doubly--periodic solutions of the
Korteweg-de Vries (KdV) equation $u_t=6uu_x-u_{xxx}$. Due to the
Moser--Trubowitz isomorphism, we get a natural relation between
the classification of real doubly--periodic geodesics and the
relevant class of periodic potentials associated to the Hill
operator $-\partial_x^2+u(x,t)$, depending on the parameter $t$
(due to the impossibility of citing all relevant contributions in
this field we limit to cite \cite{Dub_Nov,Dub,IM,L,MKVM2,AMM77}).

\smallskip

We recall that a solution to the KdV equation of the form
$u(x,t)=2\sum\limits_{j=1}^N {\cal P} (x-q_i(t))+c$ is called a
KdV--elliptic soliton. $u(x,t)$ is a KdV-elliptic soliton if and
only if $\sum\limits_{1\le j\le N, j\not =k} {\cal P}^{\prime}
(q_j(t)-q_k(t)) =0$, $k=1,\dots,N$ \cite{AMM77}. Any
KdV--elliptic soliton is uniquely associated with a marked
hyperelliptic curve $(X,P)$ of positive genus $g$ equipped with a
projection $\pi \, : X\mapsto {\cal E}$ the so called
hyperelliptic tangential cover - such that $P$ is a smooth
Weierstrass point of $X$ and the canonical images of $(X,P)$ and
$({\cal E}, Q)$ in the Jacobian of $X$ are tangent at the origin
\cite{TV2}.

The problem of classifying all hyperelliptic tangential covers in
the complex moduli space of genus $g$ hyperelliptic curves and to
characterize the associated KdV--elliptic solitons has been
successfully considered in a series of papers by Treibich and
Verdier \cite{TV}-\cite{Ver}. We refer to \cite{Tr} for an account
of the vast literature on the subject.

In particular, a different approach to the classification problem
of KdV--elliptic solitons has been developed by Krichever
\cite{Krich} based on the theory of one point Baker--Akhiezer
functions, while Gesztesy and Weikard \cite{GesWei} give an
analytic characterization of elliptic finite--gap potentials.
Finally, explicit examples of families of such coverings have been
worked out by many authors (see in particular
\cite{Smirnov1,TV,Smirnov2}).

\medskip

In \cite{AF,Fed05}, hyperelliptic tangential covers were first
considered in connection to doubly--periodic closed geodesics on
$n$--dimensional (complex) quadrics and explicit examples were
worked out. In particular, a theorem by Fedorov\cite{Fed05}
implies if the curve $(\Gamma,P_0)$ is a (complex) hyperelliptic
tangential cover, then the geodesics on the associated quadric are
(complex) doubly periodic.

\medskip

Here we restrict ourselves to real hyperelliptic curves $\Gamma$
with all finite branch points real. For such curves we call the
hyperelliptic tangential covering real (resp. real rectangular,
real rhombic), if the elliptic curve ${\cal E}$ has a real
structure (resp. with rectangular, rhombic period lattice).

The theorem by Fedorov may be easily rephrased so to hold in the
case of real tangential coverings. Moreover, here we prove the
reverse statement: if the double--periodicity condition
(\ref{realp1}) and (\ref{imp}) hold, then the associated algebraic
curve is a real hyperelliptic tangential cover. Finally, in the
latter case we show that it is always possible to associate to the
hyperelliptic curve for which the double periodicity condition
holds, a real rectangular hyperelliptic tangential covering.

The conclusion is then that the double--periodicity condition is
necessary and sufficient for the algebraic characterization of the
closedness property of real geodesics on $n$--dimensional
ellipsoids.

\begin{definition}\label{setting2} {\bf Real rectangular hyperelliptic tangential coverings}
Let $\Gamma\, : \{ \mu^2 = -\prod\limits_{k=0}^{2n}
(\lambda-b_k)\}$ be as in (\ref{gamma}), let $P_0=(0,0)$ and let
$\omega_k= \lambda^{k-1}/\mu d\lambda$, $k=1,\dots,n$, be the
basis of holomorphic differentials introduced in (\ref{basis}).
Let $A=2\left(\sqrt{-\prod_{j=1}^{2n}b_j}\right)^{-1}$ be as in
the proof of Lemma \ref{lemmavan}.

The curve $\Gamma$ admits a canonical embedding into its Jacobian
variety ${\rm Jac}(\Gamma)$ by the map $P\mapsto {\cal A} (P) =
\int_{P_0}^P (\omega_1,\dots,\omega_n)^T,$ so that $P_0$ is mapped
into the origin of the Jacobian and ${\bf U} = \left.
\frac{d}{d\tau} {\cal A} (P)\right|_{P=P_0} = ( A,0,\dots,0),$ is
the tangent vector of $\Gamma\subset {\rm Jac} (\Gamma)$ at the
origin.

Assume that $\Gamma$ is an $N$--fold covering of an elliptic curve
${\cal E}$, which we represent in the canonical Weierstrass form
\[
{\cal E} \, = \{ ({\cal P}^{\prime} (u))^2 = 4 {\cal P}^3 (u) -g_2
{\cal P} (u) -g_3 \equiv 4({\cal P} (u) -e_1)({\cal P} (u)
-e_2)({\cal P} (u) -e_3)\}.
\]
Assume that under the covering  map $\pi \, : {\Gamma} \mapsto
{\cal E}$, $P_0$ is mapped to $Q_0$ the infinite point of ${\cal
E}$ and choose $u$ as local coordinate.

The covering from the marked curve $(\Gamma, P_0)$ to $({\cal E},
Q_0)$ is hyperelliptically tangential if ${\cal E}$ admits the
following canonical embedding to ${\rm Jac}(\Gamma)$, $\; u\mapsto
u{\bf U}$, so that the embedding of $\Gamma$ and ${\cal E}$ are
tangent at the origin.

We call $(\Gamma, P_0)$ a {\bf real hyperelliptic tangential
covering} if the above holds and the elliptic curve ${\cal E}$ has
a real structure ({\it i.e.} the period lattice associated to
${\cal E}$ is either rectangular or rhombic).

We call the real hyperelliptic tangential covering $(\Gamma, P_0)$
{\bf rectangular } if moreover all the finite branch points of
${\cal E}$ are real (so the lattice associated to ${\cal E}$ is
rectangular). Otherwise, we call the real hyperelliptic tangential
covering rhombic.
\end{definition}

\medskip

For the geodesic problem, the existence of a real hyperelliptic
tangential covering implies the double-periodicity condition by
the following theorem.
\begin{theorem}\label{fed}
If $(\Gamma,P_0)$ is a real hyperelliptic tangential cover, then
the associated geodesics are closed and doubly--periodic.
\end{theorem}

The above theorem was originally proven by Fedorov \cite{Fed05} in
the complex setting: indeed if $(\Gamma,P_0)$ is a hyperelliptic
tangential cover, then the complex geodesics on the quadric $Q$
satisfy a double--periodicity condition. His argument may be
easily modified so to hold in the real setting. We remark that we
get the double-periodicity condition (\ref{realp1})-(\ref{imp})
either if the real hyperelliptic tangential covering is
rectangular or rhombic.

The above theorem settles a sufficient condition for the
algebraicity of the closedness property of real geodesics on
ellipsoids. Next theorem implies that such condition is also
necessary; so that we get the complete characterization of
algebraically closed geodesics via the double--periodicity
condition.

We now prove the converse to Theorem \ref{fed}.

\begin{theorem}\label{main}
Let $\Gamma = \{\mu^2= -\lambda \prod\limits_{i=1}^{n+1} (\lambda
- a_i) \prod\limits_{k=1}^{n-1} (\lambda-c_k) \equiv - \lambda
\prod\limits_{k=1}^{2n} (\lambda -b_k)\}$ be the hyperelliptic
curve associated to the geodesics on the ellipsoid $Q= \{
X_1^2/a_1+\cdots + X^2_{n+1}/a_{n+1} =1\}$ with constants of
motion $c_1,\dots, c_{n-1}$. Let $\alpha_1,\dots,\alpha_n$,
$\beta_1,\dots,\beta_n$ be the conventional canonical homological
basis depicted in Figure 1 and let $\omega_j$, $j=1,\dots,n$ be
the basis of holomorphic differentials introduced in
(\ref{basis}).

If the doubly--periodicity conditions (\ref{realp1}) and
(\ref{imp}) hold, then $(\Gamma, P_0)$ is a real rectangular
hyperelliptic tangential cover.
\end{theorem}

{\it Proof of Theorem \ref{main}} The doubly--periodicity
conditions (\ref{realp1}) and (\ref{imp}) hold if and only if
there exist two cycles $\alpha=\sum\limits_{i=1}^nm_i \alpha_i$
and $\beta=\sum\limits_{i=1}^n m_i^{\prime} \beta_i$, such that
\begin{equation}\label{algp}
\displaystyle \displaystyle  \oint_{\alpha} \omega_j = \left\{
\begin{array}{l} T, \quad j=1,\\
0,\quad j=2,\dots,n,\end{array}\right. \quad\quad\oint_{\beta}
\omega_j = \left\{
\begin{array}{l} \sqrt{-1}T^{\prime}, \quad j=1,
\\
\; 0, \quad\quad\quad j=2,\dots,n.
\end{array}
\right.
\end{equation}
The above equations imply that $\omega_2,\dots,\omega_n$ are the
$(n-1)$ independent holomorphic differentials vanishing at
$P_0=(0,0)$ and possess a maximal system of $(2n-2)$ independent
periods. Then by Poincar\'e reducibility theorem \cite{Poi}, there
exist an elliptic curve ${\cal E}$ and a $(n-1)$--dimensional
Abelian subvariety ${\cal A}_{n-1}$ such that ${\rm Jac} (\Gamma)$
is isogenous to the direct product ${\cal E}\times {\cal
A}_{n-1}$. Since $P_0=(0,0)$ is among the Weierstrass points of
$\Gamma$, the covering $\pi \, :\Gamma \mapsto {\cal E}$ is
tangent at the Weierstrass point $P_0$ \cite{Ver,TV2}.

Since all of the Weierstrass points of the curve $\Gamma$ are real
(see \ref{gamma}) and since the double periodicity condition
(\ref{algp}) ensures the rational dependence between the real
periods (associated to the $\alpha$ cycle) and the rational
dependence between the imaginary periods (associated to $\beta$),
we easily conclude that the hyperelliptic tangential covering has
a real structure.

We now explicitly construct such covering in order to investigate
the real structure associated to ${\cal E}$. The tangency
condition and the (\ref{algp}) ensure the existence of two real
numbers $A,B$, of a holomorphic differential $\Omega_1 = \omega_1
+\sum_{j=2}^n c_j \omega_j$, and of constants
$k_1\dots,k_n,h_1,\dots,h_n\in {\mathbb Z}$, such that
\[
\oint_{\alpha_j} \Omega_1 =2 k_j A, \quad\quad \oint_{\beta_j}
\Omega_1 = 2h_j \sqrt{-1}B,\quad j=1,\dots, n.
\]
Since $\alpha_1,\dots,\alpha_n,\beta_1,\dots,\beta_n$ form a
homological basis, any other period of $\Omega_1$ is an integer
combination of $2A$ and $2\sqrt{-1}B$. In particular,
\[
T= \oint_{\alpha} \Omega_1 = A\sum_{j=1}^n k_jm_j,\quad\quad
\sqrt{-1}T^{\prime}= \oint_{\beta} \Omega_1 =\sqrt{-1}
B\sum_{j=1}^n h_jm_j^{\prime}.\] We now investigate the real
structure of the covering. Let us fix $P_0=(0,0)\in \Gamma$ as
basepoint, let $z = \int_{P_0}^P \Omega_1$, $P\in \Gamma$. Then
$z\in {\cal T} = {\mathbb C} / \Lambda$, the one--dimensional
torus with period lattice $\Lambda$ generated by $2A, 2\sqrt{-1}
B$.

Finally let ${\cal P} (z) \equiv{\cal P} (z | A, \sqrt{-1}B)$ be
the Weierstrass ${\cal P}$-function with half-periods $A,
\sqrt{-1}B$ and ${\cal E} : \, \left\{ \big( {\cal P}^{\prime} (z)
\big)^2 = 4 \prod\limits_{k=1}^3 \big( {\cal P} (z) -e_k \big)
\right\}$ the elliptic curve in Weierstrass normal form with
finite branch points $e_1 = {\cal P} (A)$, $e_2 = {\cal P}
(A+\sqrt{-1} B)$ and $e_3 = {\cal P} (\sqrt{-1}B)$.

Then, the covering $\pi \, : \Gamma \mapsto {\cal E}$ is real
rectangular and tangential at $P_0=(0,0)$ by construction.
$\quad\quad\square$

We remark that there is a certain freedom in the construction of
the curve ${\cal E}$ and of the covering, due to the isogeneity
between ${\rm Jac} (\Gamma)$ and ${\cal E}\times {\cal A}_{n-1}$.
For instance, if we introduce the complex conjugate numbers
$C_{\pm}= A\pm \sqrt{-1}B$, we may associate to $(\Gamma,P_0)$ a
real rhombic hyperelliptic tangential covering.

\smallskip

Theorem \ref{main} means that the double-periodicity condition is
algebraic in the parameters of the problem (the square semiaxes
$a_1,\dots,a_{n+1}$ and the constants of motion
$c_1,\dots,c_{n-1}$), since it may be equivalently expressed in
terms of elliptic integrals associated to the covering.

Theorem \ref{fed} means that for the special class of geodesics on
ellipsoids associated to a real hyperelliptic tangential covering
$(\Gamma,P_0)$, the periodicity condition (\ref{realp1})
$$\sum_{i=1}^{n}  m_i \oint_{\alpha_i} \omega_1 =2
T,\quad\quad\sum_{i=1}^{n}  m_i \oint_{\alpha_i} \omega_j =0,
\;\;j=2,\dots, n,$$ is algebraic in the parameters of the problem,
since the covering $\pi$ imposes algebraic relations among the
branch points of ${\cal E}$ and the ramifications points of
$\Gamma$ (square semiaxes $a_1,\dots,a_{n+1}$ and constants of
motion $c_1,\dots,c_{n-1}$),  and the real (resp. imaginary)
periodicity condition is expressible as a real (resp. imaginary)
elliptic integral on ${\cal E}$.

We thus get the following

\begin{corollary}\label{cor}
Let $\Gamma = \{\mu^2= -\lambda \prod\limits_{i=1}^{n+1} (\lambda
- a_i) \prod\limits_{k=1}^{n-1} (\lambda-c_k) \equiv - \lambda
\prod\limits_{k=1}^{2n} (\lambda -b_k)\}$ be the hyperelliptic
curve associated to the geodesics on the ellipsoid $Q= \{
X_1^2/a_1+\cdots + X^2_{n+1}/a_{n+1} =1\}$ with constants of
motion $c_1,\dots, c_{n-1}$. Let $\alpha_1,\dots,\alpha_n$,
$\beta_1,\dots,\beta_n$ be the conventional canonical homological
basis depicted in Figure 1 and let
$\omega_j=\lambda^{j-1}d\lambda/\mu$, $j=1,\dots,n$ be the basis
of holomorphic differentials introduced in (\ref{basis}). Let
$P_0=(0,0)\in \Gamma$.

Then the closedness property (\ref{realp})
\[
\sum_{i=1}^n m_i \oint_{\alpha_i} \omega_1 =2 T,\quad\quad
\sum_{i=1}^n m_i \oint_{\alpha_i} \omega_j =0, \quad j=2,\dots, n,
\]
is algebraic in the parameters of the problem $a_1\dots,a_{n+1}$,
$c_1,\dots,c_{n-1}$ (square semiaxes and constants of motion), if
and only if there exists a non trivial imaginary cycle $\beta =
\sum_{i=1}^n m^{\prime}_i \beta_i$, such that
\[
\oint_{\beta} \omega_1 =2 \sqrt{-1}T^{\prime},\quad\quad
\oint_{\beta} \omega_j =0, \quad j=2,\dots, n.
\]
In the latter case, $(\Gamma, P_0)$ is a real rectangular
hyperelliptic tangential cover.
\end{corollary}

The Corollary is perfectly consistent with the Treibich--Verdier
characterization of elliptic solitons of the Korteweg--de Vries
equations (we refer in particular to \cite{Ver} for a discussion
of the dimension of the real moduli space associated to either the
periodic or double--periodic stationary solution to the KdV
equation).

On the other side the Corollary implies that the periodicity
condition for the geodesic problem will stay transcendental for
any over type of covering: for instance the periodicity condition
will stay transcendental, if $\Gamma$ as in (\ref{gamma}) is a
hyperelliptic tangential cover with marked point $P_j = (b_j,0)$,
for some $j=1,\dots,2n$ or if $\Gamma$ is a degree $d=2$ covering
(the degree of a hyperelliptic tangential cover is at least 3). In
particular, in the last section we prove the existence of
doubly--periodic closed geodesics related to degree 2 coverings
with extra automorphisms and we give an explicit example (see
Figure 4): in view of Corollary \ref{cor} in such case the curve
admits also a hyperelliptic tangential cover, and then an infinite
number of coverings by a classical theorem by Picard \cite{Pic}.

\medskip

\section{Density of doubly--periodic closed geodesics}
In this section, we prove that the algebraic condition of real
closed geodesics settled in the previous section, is fulfilled on
a dense set of parameters (the square semiaxes $a_1,\dots,a_{n+1}$
and the constants of motion $c_1,\dots,c_{n-1}$) with respect to
the natural topology over the reals. So it is possible to
characterize algebraically dense sets of real closed geodesics on
ellipsoids and to approximate real closed geodesics on given
ellipsoid by sequences of algebraically closed (i.e.
doubly-periodic) geodesics on perturbed ellipsoids with perturbed
constants of motion.

\smallskip

We remark that, such approximate algebraically closed geodesics
may be chosen so to share the same length and/or the same value of
the period vector as the limiting geodesic.

Our estimates are optimal in the sense that we are able to count
the number of parameters which may be kept fixed in this
approximation scheme. For instance, in the simplest case
(geodesics on triaxial ellipsoids), Theorem \ref{optimal1} implies
that we may keep fixed one parameter: indeed we have four
parameters (the three semiaxes and the caustic parameter), two
conditions originating from the limiting closed geodesics (length
$T$ and period mapping $m_1/m_2$) and one extra condition (the
approximating geodesics have rational value of the imaginary
period mapping $m_1^{\prime}/m_2^{\prime}$ which approximates the
irrational quasi--period of the limiting imaginary geodesic).
Similarly Theorem \ref{optimal2} implies that we may keep fixed
two parameters (since we also perturb the length of the
approximating algebraic geodesics).

The proofs of the density results rely on a theorem by McKean and
van Moerbeke for the Hill problem \cite{MKVM}. Using their idea,
we define a quasi-period vector $(x,y)\equiv
(x_1,\dots,x_n,y_1,\dots,y_n)\in {\mathbb R}^{2n}$ associated to
any real and imaginary geodesics. Using the Riemann bilinear
relations, such quasi--period vector may be explicitly computed
using the periods of two meromorphic differentials. The theorem by
\cite{MKVM} (originally stated for the Hill problem), ensures that
the map from the parameter space ($a_1,\dots,a_{n+1}$,
$c_1,\dots,c_{n-1}$) to the quasi periods $(x,y)$ is analytic and
locally invertible.

\paragraph{Density of algebraically closed geodesics}
For an easier comparison with the density characterization of
KdV-elliptic solitons, we also report the following
characterization of hyperelliptic tangential covers in the complex
moduli space of hyperelliptic curves due to Colombo {\it et al.}
\cite{CPP}. Their theorem implies immediately that real closed
geodesics may be approximated by complex doubly--periodic
geodesics.

\begin{theorem} \cite{CPP} Hyperelliptic tangential covers of genus $n$
are dense in the complex moduli space ${\cal H}_n$ of the
hyperelliptic curves of genus $n$. \end{theorem}

To prove the density statement (Theorem \ref{DD}) for real doubly
periodic geodesics on $n$--dimensional ellipsoids with respect to
the real parameter space, we apply the ideas used by McKean and
VanMoerbeke in \cite{MKVM} for the Hill problem. We report their
theorem below in a version suitable for the geodesics problem and
then show that any real closed geodesics on a given ellipsoid may
be approximated by real doubly--periodic geodesics on perturbed
ellipsoids.

\begin{theorem}\label{MM} \cite{MKVM}
Let $\Gamma = \{\mu^2= -\lambda \prod\limits_{i=1}^{n+1} (\lambda
- a_i) \prod\limits_{k=1}^{n-1} (\lambda-c_k) \equiv - \lambda
\prod\limits_{k=1}^{2n} (\lambda -b_k)\}$ be as in (\ref{gamma}).
Let $(x,y)=(x_1,\dots,x_n,y_1,\dots,y_n)\in {\mathbb R}^{2n}$ be
defined by
\begin{equation}\label{MMcond}
\begin{array}{l}\displaystyle \sum_{i=1}^n x_i \oint_{\alpha_i} \omega_j = \left\{
\begin{array}{lcl} 1, &{\rm for} &i=1,\\
0, &{\rm for} &i=2,\dots,n,
\end{array}\right.
\\
\displaystyle \sum_{i=1}^n y_i \oint_{\beta_i} \omega_j = \left\{
\begin{array}{lcl} \sqrt{-1}, &{\rm for} &i=1,\\
0, &{\rm for} &i=2,\dots,n.
\end{array}\right.
\end{array}
\end{equation}
Then, (\ref{MMcond}) define a real analytic locally invertible map
from open sets in the parameter space $(b_1,\dots,b_{2n})$ to open
sets in the quasi--period space $(x,y)=(x_1, \dots ,x_n, y_1,$
$\dots, y_n)$. In particular, a small perturbation of the real
branch points of $\Gamma$ will make the point $(x,y)$ rational.
\end{theorem}

If we compare (\ref{MMcond}) with the double-periodicity condition
settled in (\ref{realp1}) and (\ref{imp}),
\[
\begin{array}{l}\displaystyle \sum_{i=1}^n m_i \oint_{\alpha_i} \omega_j = \left\{
\begin{array}{lcl} T, &{\rm for} &i=1,\\
0, &{\rm for} &i=2,\dots,n,
\end{array}\right.
\\
\displaystyle \sum_{i=1}^n m_i^{\prime} \oint_{\beta_i} \omega_j =
\left\{
\begin{array}{lcl} \sqrt{-1} T^{\prime}, &{\rm for} &i=1,\\
0, &{\rm for} &i=2,\dots,n,
\end{array}\right.
\end{array}
\]
we easily conclude that if the point $(x,y)$ is rational, then the
double periodicity condition is satisfied. Then the following
density property of algebraically closed geodesics holds.

\begin{theorem}\label{DD}
Given a real closed geodesic on the ellipsoid $Q= \{ X_1^2/a_1 +
\cdots +X_{n+1}^2/a_{n+1} =1\}$ with caustic parameters $c_j$,
$j=1,\dots,n-1$, for any $\epsilon>0$ sufficiently small, there
exist
$a_1^{\epsilon},\dots,a_{n+1}^{\epsilon},c_1^{\epsilon},\dots,c_{n-1}^{\epsilon}\in
{\mathbb R}$ such that
\[\sum\limits_{j=1}^{n-1}(c_j-c_j^{\epsilon})^2+\sum\limits_{i=1}^{n+1}(a_i
-a_i^{\epsilon})^2<\epsilon\] and the geodesics on $Q^{\epsilon}=
\{ X_1^2/a_1^{\epsilon}+ \cdots +X_{(n+1)}^2/a_{n+1}^{\epsilon}
=1\}$ with caustic parameters $c_j^{\epsilon} $, $j=1,\dots,n-1$,
are real doubly periodic.
\end{theorem}

{\it Proof:} Let $\Gamma$ be the real Hill curve associated to the
closed geodesics on the ellipsoid $Q$ with caustic parameters
$c_1,\dots,c_{n-1}$ so that the set of equations (\ref{realp1})
hold. Let $\epsilon_0 = \frac{1}{2} \min \{ b_j -b_{j-1}, \;
j=1,\dots,2n \}$ where $\{b_1<\dots<b_{2n}\} =
\{a_1<\dots<a_{n+1}, c_1<\cdots,c_{n-1}\}$.

$\Gamma^{\prime}$, the dual curve to $\Gamma$ introduced in
Theorem \label{lemma3}, is associated to a dual ellipsoid
$Q^{\prime}$ which possesses real quasi--periodic closed
geodesics, so that the vector $y=(y_1,\dots,y_n)\in {\bf R}^n$.

Similarly to \cite{MKVM}, we introduce the differential of the
second kind $\Omega^{(0)}_2$ with vanishing $\beta_j$ periods,
with a double pole at $P_0=(0,0)$ and the following normalization.
Let $\tau$ be the local coordinate in a neighborhood of
$P_0=(0,0)$ such that $\tau(P_0)=0$, then $\Omega^{(0)}_2 \approx
(2\pi A)^{-1}\tau^{-2}d\tau$, $k=1,\dots,n$, where
$A=2\left(\sqrt{-\prod_{j=1}^{2n}b_j}\right)^{-1}$ is the constant
defined in Definition \ref{setting2}. Let
\[
y_j = \oint_{\alpha_j} \Omega^{(0)}_2, \quad j=1,\dots, n,
\]
be the $\alpha$--period vector of $\Omega^{(0)}_2$. Then applying
Riemann bi--linear identities to $\omega_l$, $l=2,\dots,n$ and to
${\tilde u} = \int_{P_0}^P \Omega^{(0)}_2$, we immediately
conclude that
\[
\sum_{j=1}^n y_j \oint_{\beta_j} \omega_1 =\sqrt{-1},\quad\quad
\sum_{j=1}^n y_j \oint_{\beta_j} \omega_l =0, \quad\quad
l=2,\dots,n.
\]
Finally, applying Theorem \ref{MM}, we may perturb the curve
$\Gamma$ so that on $\Gamma^{\epsilon}$ (with
$\epsilon<\epsilon_0$)
\[
\sum_{j=1}^n x_j^{(\epsilon)} \oint_{\alpha_j^{(\epsilon)}}
\omega_1^{(\epsilon)} =1, \quad\quad \sum_{j=1}^n y_j^{(\epsilon)}
\oint_{\beta_j^{(\epsilon)}} \omega_1^{(\epsilon)} =\sqrt{-1},
\]
\[
\sum_{j=1}^n x_j^{(\epsilon)} \oint_{\alpha_j^{(\epsilon)}}
\omega_l^{(\epsilon)} =0, \quad\quad \sum_{j=1}^n y_j^{(\epsilon)}
\oint_{\beta_j^{(\epsilon)}} \omega_l^{(\epsilon)} =0,\quad\quad
l=2,\dots,n,
\]
for rational vector $(x^{(\epsilon)},y^{(\epsilon)})$. According
to Theorem \ref{main} $(\Gamma^{(\epsilon)},P_0)$ is a
hyperelliptic tangential cover.
 $\quad \square$

\medskip

\begin{remark}{\rm
It is easy to verify that the vectors $(x,y)$ in (\ref{MMcond})
correspond to a hyperelliptic tangential cover if and only if
$x=(x_1,\dots,x_n)$ has rationally dependent components and the
same holds for $y=(y_1,\dots,y_n)$ (that is the requirement that
$(x,y)$ be rational may be weakened, without loosing the
algebricity of the closedness condition of the associated
geodesics)}.
\end{remark}

In view of the above remark, it is possible to optimize the
density characterization of doubly periodic closed geodesics.
Indeed it is possible to modify the proof of the above theorem so
 that the ellipsoids $Q$, $Q^{\epsilon}$ share the same value of
the greatest square semiaxis $a_{n+1}=a_{n+1}^{\epsilon}$, and the
perturbed real doubly-periodic closed geodesics on $Q^{\epsilon}$
have the same length and the same period vector as the initial
real closed geodesics on $Q$, {\it i.e.} $(x_1,\dots,x_n)=
(x_1^{\epsilon},\dots,x_n^{\epsilon})$.

\begin{theorem}\label{optimal1}
Let $\Gamma = \{ \mu^2 =
-\lambda(\lambda-c)\prod\limits_{i=1}^{n+1} (\lambda
-a_i)\prod\limits_{j=1}^{n-1}(\lambda-c_j) \}$, be a real Hill
curve so that the real geodesics on the ellipsoid $Q= \{ X_1^2/a_1
+ \cdots +X_{n+1}^2/a_{n+1} =1\}$ with caustic parameters
$c_1,\dots,c_{n-1}$ are closed and have length $T$.

Then, there exists a sequence $\{a_1^{(k)},\dots,
a_n^{(k)},c_1^{(k)},\dots,c_{n-1}^{(k)}\}\in {\mathbb R}^{2n-1}$
such that
\[\lim\limits_{k\to +\infty} c_j^{(k)} =c_j, \; (j=1,\dots,n-1), \quad\quad
\lim\limits_{k\to +\infty} a_i^{(k)} =a_i, \; (i=1,\dots, n),\]
and the geodesics on $Q^{(k)}= \{ X_1^2/a_1^{(k)} + \cdots
+X_{n}^2/a_{n}^{(k)}+X_{n+1}^2/a_{n+1} =1\}$ with caustic
parameters $c^{(k)}=(c_1^{(k)},\dots,c_{n-1}^{(k)})$ are
doubly--periodic, with same length $T$ and with the same value of
the period vector as the closed geodesics on
$(Q,c_1,\dots,c_{n-1})$.
\end{theorem}

{\it Proof:} The proof follows from a straightforward adaptation
of the argument in Theorem \ref{MM}: since the jacobian
determinant of the real analytic map there defined is not
vanishing, also its restriction to a generic $2n-1$--dimensional
subvariety will not vanish locally. To fix ideas, we choose the
subvariety $b_{2n}\equiv a_{n+1}=const.$.

Let $\Gamma$ be real Hill, let $\omega_1,\dots,\omega_n$ be the
holomorphic basis of differentials defined in (\ref{basis}) and
$\alpha_i,\beta_i$, $i=1,\dots,n$ the homological basis as in
Remark \ref{rem2}.

Let $\Omega^{(0)}_2$ be the normalized meromorphic differential of
the second kind with double pole at $P_0=(0,0)$, vanishing $\beta$
periods, as in the proof of Theorem \ref{DD}, and let
$(y_1,\dots,y_n)$ be its $\alpha$ period vector.

Let $\epsilon_0 = \frac{1}{2} \min \{ b_j -b_{j-1}, \;
j=1,\dots,2n \}$, where, as usual $\{ b_1<\cdots<b_{2n}\} = \{
a_1,\dots,a_{n+1},c_1,\dots,c_{n-1}\}$.

Then the geodesics on $Q$ with caustic parameters
$c_1,\dots,c_{n-1}$ are real closed and satisfy the periodicity
condition
\begin{equation}\label{rp}
\begin{array}{l}
\displaystyle f_1 (b_1,\dots,b_{2n}) \equiv \sum_{i=1}^n m_i
\oint_{\alpha_i} \omega_1 - T=0,\\
\displaystyle  f_j (b_1,\dots,b_{2n}) \equiv \sum_{i=1}^n m_i
\oint_{\alpha_i} \omega_j = 0, \;\quad\quad j=2,\dots,n.
\end{array}
\end{equation}
Let $b_{2n},m_1,\dots,m_n,T$ be fixed. As a consequence of Theorem
\ref{MM}, the $n$ equations $f_j=0$, $j=1,\dots,n$ are locally
analytically invertible near the point $(b_1,\dots,b_{2n-1})$ and
there exist $n$ analytic functions ${\hat b}_{r}= {\hat b}_r
({\hat b}_1,\dots, {\hat b}_{n-1})$, $r=n,\dots,2n-1$, on the
$(n-1)$--dimensional ball ${\cal B}_0$ centered at
$(b_1,\dots,b_{n-1})$ and of radius $\epsilon<\epsilon_0$.

On the initial curve $\Gamma$,
\[g_j  \equiv y_j/y_n =
\int_{b_{2j-1}}^{b_{2j}}\Omega^{(0)}_2/\int_{b_{2n-1}}^{b_{2n}}\Omega^{(0)}_2,\quad\quad
j=1,\dots,n-1\] take some real value $\tau_j$, $j=1,\dots,n-1$ and
are real analytic in ${\hat b}_1,\dots, {\hat b}_{n-1}$ on the
ball ${\cal B}_0$, again by Theorem \ref{MM}.

Then, there exists a sequence $(b_1^{(k)},\dots,b_{n-1}^{(k)}) \in
{\cal B}_0$ converging to $(b_1,\dots b_{n-1})$ such that
\begin{equation}\label{ip}
\begin{array}{ll}
\displaystyle \lim\limits_{k\to +\infty} b^{(k)}_r \equiv
\lim\limits_{k\to +\infty}{\hat b}_r
(b_1^{(k)},\dots,b_{n-1}^{(k)}) =b_r, &\quad\quad
r=n,\dots,2n-1;\\
&\\
\displaystyle \quad g_j (b_1^{(k)},\dots,b_{n-1}^{(k)})\in
{\mathbb Q},&
\quad\quad j=1,\dots,n-1; \\
&\\ \displaystyle \lim\limits_{k\to +\infty}
g_j(b_1^{(k)},\dots,b_{n-1}^{(k)}) =\tau_j, &\quad\quad
j=1,\dots,n-1.
\end{array}
\end{equation}

Finally, for any $k$, by construction, the corresponding
hyperelliptic curve $\Gamma^{(k)} = \{ \mu^2 =
-\lambda\prod\limits_{j=1}^{2n} (\lambda-b_j^{(k)})\}$ is a
hyperelliptic tangential cover with marked point $P_0=(0,0)$ and
the associated geodesics have the same length and the same period
vector as the initial ones associated to $\Gamma$. Indeed
equations (\ref{rp}) ensure that on $\Gamma^{(k)}$ the period
vector and the length of the real geodesics be preserved; by
(\ref{ip}), the imaginary period vector
$(y_1^{(k)},\dots,y_n^{(k)})$ has rationally dependent components
for all $k$ which, approximate the rationally independent
components of the imaginary quasi--period of the limiting
imaginary geodesics, so that, by construction, the limiting real
closed geodesics are those associated to $\Gamma$. $\quad\quad
\square$

\medskip

Finally, if we just require to preserve the period vector of the
geodesics and allow that the length of the approximating geodesics
vary, {\i.e.} if we just require $(x_2/x_1,\dots,x_n/x_1)=
(x_2^{\epsilon}/x_1^{\epsilon},\dots,x_n^{\epsilon}/x_1^{\epsilon})$,
we may keep fixed two square semiaxes, for instance the smallest
and the greatest one, $a_1=a_1^{\epsilon}$ and
$a_{n+1}=a_{n+1}^{\epsilon}$ and we get the following statement.

\begin{theorem}\label{optimal2}
Let $\Gamma = \{ \mu^2 =
-\lambda(\lambda-c)\prod\limits_{i=1}^{n+1} (\lambda
-a_i)\prod\limits_{j=1}^{n-1}(\lambda-c_j) \}$, be a real Hill
curve so that the real geodesics on the ellipsoid $Q= \{ X_1^2/a_1
+ \cdots +X_{n+1}^2/a_{n+1} =1\}$ with caustic parameters
$c_1,\dots,c_{n-1}$ are closed and have length $T$.

Then, there exists a sequence $\{a_2^{(k)},\dots,
a_n^{(k)},c_1^{(k)},\dots,c_{n-1}^{(k)}\}\in {\mathbb R}^{2n-2}$
such that
\[
\lim\limits_{k\to +\infty} c_j^{(k)} =c_j,\; (j=1,\dots,n-1),
\quad\quad \lim\limits_{k\to +\infty} a_i^{(k)} =a_i, \;
(i=2,\dots, n),\] and the geodesics on $Q^{(k)}= \{ X_1^2/a_1 +
X_2^2/a_2^{(k)} +\cdots +X_{n}^2/a_{n}^{(k)}+X_{n+1}^2/a_{n+1}
=1\}$ with caustic parameters
$c^{(k)}=(c_1^{(k)},\dots,c_{n-1}^{(k)})$ are doubly--periodic,
with same value of the period vector as the closed geodesics on
$(Q,c_1,\dots,c_n)$.
\end{theorem}

\medskip

The proof is a straightforward modification of the one for Theorem
\ref{optimal1} and we omit it.

\paragraph{Remark} In \cite{AF}, we used the algebraic
characterization of closed geodesics associated to hyperelliptic
tangential covers to construct periodic billiard trajectories of
an integrable billiard on a quadric $Q$ with elastic impacts on a
confocal quadric $Q_d$. The results we have presented in this
section may be applied to this billiard model and imply the
algebraic characterization of a dense set of its periodic orbits.

\section{The algebraic computation of the period mapping in the case $n=2$}

In the special case of triaxial ellipsoids ($n=2$), a stronger
characterization of doubly-periodic closed geodesics holds. In
particular, we show below that the period mapping of a doubly
periodic closed geodesic, which measures the ratio between
oscillation and winding for a geodesics, is algebraic in the
parameters of the problem and that it may be explicitly computed
using the second covering associated to the hyperelliptic curve.
Indeed, the 2--dimensional ${\rm Jac} (\Gamma)$ is isogenous to
the product of two elliptic curves ${\cal E}_1\times {\cal E}_2$.

\smallskip

The second covering plays a relevant role also in the case of
elliptic solitons. Airault {\sl et al.} \cite{AMM77} discovered a
remarkable link between the pole dynamics of the KdV elliptic
solutions with the initial data in the form of the Lam\'e
potential and the dynamics of Calogero--Moser particle system
\cite{Cal}. In the genus 2 case, the topological characterization
of the covering ramified at $P_0$ reduces the problem of
describing the pole dynamics to the search of solutions of certain
algebraic equations related to the covering and to the inversion
of elliptic integrals \cite{Enol, Smirnov2}.

Below we first recall the definition of the period mapping and
some classical results. Then we show how to compute the period
mapping explicitly using the topological character of the
associated second covering. Unfortunately there do no exist
general theorems which characterize topologically such families of
coverings. As an application, we compute the value of the period
mapping for some special classes of coverings in the next section.

\paragraph{Closed geodesics on triaxial ellipsoids and the period mapping}
In the case $n=2$ (geodesics on triaxial ellipsoids), Proposition
\ref{1} implies that for any fixed choice of the semiaxes
$0<a_1<a_2<a_3$ there is a dense set $I\subset ]a_1,a_3[\backslash
\{ a_2 \}$ such that for all $c\in I$ the hyperelliptic curve
$\Gamma \; :\, \{ \mu^2 = -\lambda (\lambda -c)
\prod\limits_{i=1}^3 (\lambda-a_i)\}$ is Hill. The application
\begin{equation}\label{permap}
c\mapsto \displaystyle \varphi(c) = \left\{
\begin{array}{ll}  2\oint_{\alpha_2} \omega_2
: 2\oint_{\alpha_1}\omega_2, &\quad a_1<c<a_2<a_3,\\
&\\ 2\oint_{\alpha_1} \omega_2 : 2\oint_{\alpha_2}\omega_2, &\quad
a_1<a_2<c<a_3,
\end{array}\right.
\end{equation}
measures the ratio between oscillation and winding for a geodesics
with parameter $c$ and it is called the period mapping (see
\cite{Klin}). Comparing the above definition with (\ref{realp1})
and Proposition \ref{1}, it is evident that the geodesic with
parameter $c$ is closed if and only if $\varphi(c)$ is rational. A
closed geodesic is called simple if it has no self-intersections.
To be simple closed, only a single winding is allowed; hence
$\varphi(c)$ must be an integer greater than one. The following
theorems explain under which condition there do exist
topologically non--trivial simple closed geodesics.

\begin{theorem} \cite{Klin}
Let $a_1<a_2<a_3$ be fixed and $c\in ]a_1,a_3[\backslash \{a_2\}$.
Then $\varphi(c)$ is a monotone decreasing function of $c$. If
$c\in ]a_1, a_2[$, then $\varphi(c)>1$ and $\lim\limits_{c\to a_2}
\varphi(c)=1$. If $c\in ]a_2,a_3[$ then $\varphi(c)<1$ and
$\lim\limits_{c\to a_2} \varphi(c)=1$.

Moreover, let $t=a_1/a_3$ be fixed and $\sigma=a_2/a_3 \in ]t,1[$.
Then, $\varphi(a_1)$ is a monotone increasing function of $\sigma$
with upper limit $\sqrt{a_3/a_1}$ and lower limit $1$.
\end{theorem}

\begin{theorem} \cite{Klin}
On an ellipsoid $\{ \sum\limits_{i=1}^3 X_i^2/a_i =1 \}$, there
exist non standard simple closed geodesics (i.e. simple closed
geodesics different from the three principal ellipses), if and
only if $\varphi(a_1)>2$.

More precisely, for each integer value $\varphi(c) \in ]1,
\varphi(a_1)[$, the projection of the flow lines yields closed
geodesics which wind once around the $X_1$--axis while performing
$\varphi(c)$ many oscillations. Their length is greater than the
length of the middle ellipse in the $(X_1,X_3)$--plane.
\end{theorem}

\medskip

\paragraph{The second covering}
In the case $n=2$, ${\rm Jac} (\Gamma)$ is isogenous to the
product of two elliptic curves ${\cal E}_1\times {\cal E}_2$ and
the second covering is ramified at $P_0=(0,0)$ according to the
following proposition by Colombo {\it et al}

\begin{proposition}\label{prop4.3} \cite{CPP}
Let $\Gamma$ be a genus 2 curve which covers an elliptic curve
$\pi_1 \, : \Gamma\mapsto {\cal E}_1$ and let $\pi_2 \, :
\Gamma\mapsto {\cal E}_2$ be another covering so that ${\rm Jac} (
\Gamma) \approx {\cal E}_1 \times {\cal E}_2$. Then $\pi_i$ is
tangential exactly at the points where $\pi_j$ is ramified $i\not
=j$.
\end{proposition}

\medskip

We briefly turn back to the double-periodicity condition in the
special case of geodesics on triaxial ellipsoids so to construct
directly the second covering associated to the double--periodicity
condition.

\begin{proposition}\label{lemmaseccov}
Let $\Gamma = \{\mu^2= -\lambda(\lambda-c) \prod\limits_{i=1}^{3}
(\lambda - a_i) \equiv - \lambda \prod\limits_{k=1}^{4} (\lambda
-b_k)\}$ be the genus 2 hyperelliptic curve associated to the
geodesics on the triaxial ellipsoid $Q= \{ X_1^2/a_1+X_2^2/a_2 +
X^2_{3}/a_{3} =1\}$ with caustic parameter $c$. Let
$\alpha_1,\alpha_2$, $\beta_1,\beta_2$ be the conventional
canonical homological basis depicted in Figure 1 and let
$\omega_1=d\lambda/\mu$, $\omega_2=\lambda d\lambda/\mu$, be the
basis of holomorphic differentials introduced in (\ref{basis}).

Suppose that on $\Gamma$ as above, the double periodicity
condition (\ref{realp1}) and (\ref{imp}) holds
\[
\begin{array}{ll}
\displaystyle m_1\oint_{\alpha_1} \omega_1 +m_2\oint_{\alpha_2}
\omega_1 =2 T, &\displaystyle \quad\quad  m_1\oint_{\alpha_1}
\omega_2 +m_2\oint_{\alpha_2} \omega_2=0,
\\
&\\ \displaystyle m_1^{\prime}\oint_{\beta_1} \omega_1
+m_2^{\prime}\oint_{\beta_2} \omega_1 =2
\sqrt{-1}T^{\prime},&\displaystyle \quad\quad
m_1^{\prime}\oint_{\beta_1} \omega_2 +m_2^{\prime}\oint_{\beta_2}
\omega_2=0, \end{array}
\] for some non--zero real $T,T^{\prime}$.
Then, there exists a covering
\[
\pi_2 \, : \; \Gamma \mapsto {\cal E}_2,
\]
ramified of order 3 at $P_0=(0,0)$ and such that
\[
\pi_2^* (\Omega_2) = \kappa \omega_2,
\]
where $\Omega_2$ is the normalized holomorphic differential on
${\cal E}_2$ and $\kappa$ is a numerical constant.
\end{proposition}

{\it Proof:} The double--periodicity conditions (\ref{realp1}) and
(\ref{imp}) imply the existence of two cycles $\alpha=m_1 \alpha_1
+m_2 \alpha_2$ and $\beta=m_1^{\prime} \beta_1 +m_2^{\prime}
\beta_2$, such that
\begin{equation}\label{algp2}
\oint_{\alpha} \omega_2 = 0,\quad \quad  \oint_{\beta} \omega_2 =
0.
\end{equation} In (\ref{algp2}) it is not restrictive to suppose
that $(m_1,m_2)$ (respectively $(m_1^{\prime}, m_2^{\prime})$), be
relative prime integer numbers.

Inspection of (\ref{algp2}) implies that all of the periods of
$\omega_2$ are integer multiples of two periods $S,
\sqrt{-1}S^{\prime}$ of $\omega_2$ and this is sufficient to prove
the existence of a covering $\pi_2 \, : \, \Gamma \mapsto {\cal
E}_2$.

Indeed, let $(m_1,m_2)$ (respectively $(m_1^{\prime},
m_2^{\prime})$), be relative prime integer numbers and let
$2S=\oint_{\alpha_1} \omega_2/|m_2|$ (resp.
$2\sqrt{-1}S^{\prime}=\oint_{\beta_1} \omega_2/|m_2^{\prime}|$).
By Bezout identity, there exist integers $p_j, p^{\prime}_j$,
$j=1,2$ such that $p_1m_1-p_2m_2=1$ (resp.
$p_1^{\prime}m_1^{\prime}-p_2^{\prime}m_2^{\prime}=1$ so that $2S,
2\sqrt{-1}S^{\prime}$ are indeed periods of $\omega_2$ and any
other period $\oint_{\gamma} \omega_2$ is an integer multiple of
$2S, 2\sqrt{-1}S^{\prime}$.

Let now fix $P_0=(0,0)\in \Gamma$ as basepoint, let $z =
\int_{P_0}^P \omega_2$, $P\in \Gamma$. Then, by Poincar\'e
reducibility theorem, $z\in {\cal T} = {\mathbb C} / \Lambda$, the
one--dimensional torus with period lattice $\Lambda$ generated by
$2S, 2\sqrt{-1} S^{\prime}$.

Finally let ${\cal P} (z) \equiv{\cal P} (z | S,
\sqrt{-1}S^{\prime})$ be the Weierstrass ${\cal P}$-function with
half-periods $S, \sqrt{-1}S$ and ${\cal E}_2 : \, \left\{ \big(
{\cal P}^{\prime} (z) \big)^2 = 4 \prod\limits_{k=1}^3 \big( {\cal
P} (z) -E_k \big) \right\}$ the elliptic curve in Weierstrass
normal form with finite branch points $E_1 = {\cal P} (S)$, $E_2 =
{\cal P} (S+\sqrt{-1} S^{\prime})$ and $E_3 = {\cal P}
(\sqrt{-1}S^{\prime})$.

Then, the covering $\pi_2 \, : \Gamma \mapsto {\cal E}_2$ has
degree $d$ and, introducing local coordinates at $P_0\in \Gamma$,
it is straightforward to verify that it is ramified of order three
at $P_0=(0,0)$ (the latter remark implies $d\ge 3$). $\quad\quad
\square$

\medskip

Of course, by Theorem \ref{main}, we already know that there
exists a covering $\pi_1 \, : \Gamma \mapsto {\cal E}_1$ which is
hyperelliptic tangential at $P_0$. The second covering constructed
above is ramified exactly at $P_0$ in agreement with Proposition
\ref{prop4.3}.

\medskip

\paragraph{The second covering and the period mapping}
We now show that the topological type of the second covering
(which is ramified at $P_0$ of order 3) is naturally linked to the
topological classification of the associated real closed geodesics
(period mapping).

\begin{definition}{\bf (topological characteristic of the second
covering)} The topological characteristic of a covering is a
sequence of four integer numbers $(\nu_0, \nu_1,\nu_2,\nu_3)$
which count the number of Weierstrass points of $\Gamma $ in the
preimage of the four branch point of ${\cal E}_2$, with the
exception of $P_0=(0,0)\in {\Gamma}$, the Weierstrass point at
which the second covering is ramified, and with the usual
convention that $\nu_0$ is associated to the branch point of
${\cal E}_2$ at infinity.
\end{definition}

For a given $\Gamma = \{ \mu^2 = -\lambda \prod\limits_{j=1}^4
(\lambda-b_j)\}$, the computation of the period mapping amounts to
identify the two integer numbers $m_1,m_2$ such that
\[
m_1 \oint_{\alpha_1}\omega_2 + m_2 \oint_{\alpha_2} \omega_2 =0.
\]
Let $\pi_2\, :\, \Gamma\to {\cal E}_2$ be the second covering,
where ${\cal E}_2 = \{ {\cal W}^2= 4{\cal Z}^3 -G_2 {\cal Z}-G_3
\equiv 4\prod\limits_{i=1}^3 (Z-E_i) \}$ is represented in the
canonical Weierstrass form. In our setting  the curves $\Gamma$
and ${\cal E}$ are real with maximal number of real connected
components, so that it makes sense to call $\alpha$ the real cycle
associated to ${\cal E}$.

From the proof of Proposition \ref{lemmaseccov}, we know that the
pull-back of the holomorphic differential on ${\cal E}_2$ is $dZ/W
= \omega_2$. So we may conclude that
\begin{equation}\label{percov} \oint_{\alpha_i} \omega_2 = \kappa_i
\oint_{\alpha} {dZ}/W, \quad i=1,2, \end{equation} where the
integer numbers $\kappa_1,\kappa_2$ satisfy
$m_1\kappa_1+m_2\kappa_2=0$.

Finally, $\kappa_1$, $\kappa_2$ are uniquely associated to the
topological characteristic of the covering $\pi_2$. To compute
them it is sufficient to compute the preimages of the four branch
points $E_0,E_1,E_2,E_3$ of ${\cal E}_2$. Since in our setting the
covering is real, $\pi_2^{-1} (E_i)$ are either the branch points
of $\Gamma$ or real points on the curve $\Gamma$ or come in
complex conjugate pairs.

Then it is self--evident that, whenever we know the topological
characteristic of second covering, we may compute $\kappa_1$ and
$\kappa_2$. Unfortunately, we do not possess such complete piece
of information in the general case. Anyway, for any degree $d$,
there exist a finite number of families of hyperelliptic
tangential coverings so that only a finite number of topological
characteristic are possible and, consequently, only a finite
number of values of the period mapping may be realized. In the
next section we discuss the case in which the degree of the
covering is either 3 or 4. When the degree of the covering is $5$,
there exist two families of hyperelliptic tangential coverings
(see \cite{Smirnov2}) and there exist real doubly--periodic
geodesics associated to such coverings either simple or with 1,2,3
or 4 self--intersections. Since the complexity of the computations
increases with the degree of the covering, we shall report the
degree $d=5$ case in detail in a subsequent publication.

\section{Examples and applications}
Explicit examples of hyperelliptic tangential covers when the
genus of the hyperelliptic curve is $n\le 8$ have been worked out
(see for instance \cite{Tr} and references therein).

In this section, we impose the reality conditions for algebraic
closed geodesics on triaxial ellipsoids for the families of degree
$d=3,4$ hyperelliptic tangential covers and we determine the
possible values of the period mapping using the topological
character of the second covering. For a comparison with the case
of elliptic KdV solitons, we refer to Smirnov \cite{Smirnov1,
Smirnov2} or to  Belokolos and Enol'ski \cite{Enol}.

Finally in the last subsection, we prove the existence of
doubly--periodic closed geodesics related to degree 2 coverings
with extra automorphisms and we give an explicit example (see
Figure 4): in view of Corollary \ref{cor} in such case the curve
admits also a hyperelliptic tangential cover, and then an infinite
number of coverings by a classical theorem by Picard \cite{Pic}.
The same family of coverings has also been considered by Taimanov
\cite{Tai} in relation to elliptic KdV solitons.

\begin{remark}
In all examples, we adopt the following convention:
$0<a_1<a_2<a_3$ are the semiaxes of the triaxial ellipsoid
$Q=\{X_1^2/a_1+X_2^2/a_2+X_3^2/a_3=1\}$ and $c$ is the parameter
of the confocal quadric to which the geodesic is tangent, so that
the finite branch points of the associated hyperelliptic curve
$\Gamma$ are $\{b_0=0<b_1<b_2<b_3<b_4\} =\{ 0, c, a_i,
i=1,\dots,3\}$.
\end{remark}

For an easier comparison of our results with $d$-elliptic KdV
solitons, we first impose that the hyperelliptic tangential $d:1$
covering $({\cal G}, P_{\infty} )\mapsto ({\cal E},Q)$ be
associated to real KdV-solitons, where ${\cal G} \, : \; \{ w^2 =
-\prod_{k=1}^5 (z-z_k) \}$ and $P_{\infty}$ is the branch point of
${\cal G}$ at infinity. Then, by Moser--Trubowitz isomorphism, the
curves ${\cal G}$ and $\Gamma$ are birationally equivalent and the
following relation among the finite branch points $z_k$s of ${\cal
G}$ and the finite branch points $b_j$ of $\Gamma$ holds:
\begin{equation}\label{zetafs} \{ z_1,z_2,z_3,z_4,z_5\} = \{
\beta, \beta+\frac{1}{b_j},\; j=1,\dots, 4\},\quad {\rm where} \;
\beta=\min \{ z_k,\; k=1,\dots,5\}.\end{equation}

\subsection{Hyperelliptic tangential covers of degree $3$.}

\paragraph{Description of the hyperelliptic tangential covering and reality problem
for doubly--periodic closed geodesics} The tangential 3:1 covering
${\cal G}\mapsto {\cal E}$ is associated to 3-elliptic KdV
solutions and dates back to the works of Hermite and Halphen
(\cite{Herm}). For the closed geodesics problem, we require that
the genus 2 curve ${\Gamma}$ is birationally equivalent to ${\cal
G} =\left\{w^2 = -\frac 14 (4 z^3-9 g_2 z- 27 g_3 ) (z^2 -3g_2)
\right \}$ which covers the elliptic curve ${\cal E}_1 =\{  W^2 =
4 Z^3 - g_2 Z -g_3 \}$, where the covering is given by the
relations
\[
\displaystyle Z=-\frac 19 \frac{z^3-27 g_3}{z^2-3g_2},
\quad\quad\displaystyle W=\frac 2{27} \frac {w(z^3 -9g_2 z+ 54
g_3)}{(z^2-3g_2)^2}.
\]
The holomorphic differential on ${\cal
E}_1$ is the pull--back of the holomorphic differential
$\displaystyle\frac {d Z}{W}= -\frac{3}{2}\frac {z\, d z}{w}$on
${\cal G}$.

The $z_j$s are related to the branch points of ${\cal E}_1$,
$e_j$, $j=1,\dots,3,$ by
\[
\beta\equiv z_1=-\sqrt{3g_2}\; <\;z_2=3e_1\;<\;
z_3=3e_2\;<\;z_4=3e_3\;<\;z_5=\sqrt{3g_2}, \] where $e_1<e_2<e_3$.

\begin{proposition} \label{e1-e4}
Let $0<a_1<a_2<a_3$ and $c\in ]a_1,a_3[\backslash \{a_2\}$ be
given. Then the geodesic flow on the ellipsoid $Q$  tangent to the
confocal quadric $Q_c$ is doubly periodic and related, up to
birational transformation, to the 3:1 covering ${\cal G}\to {\cal
E}_1$ if and only if
\begin{equation}\label{condA}
\frac{1}{c^2}+\frac{1}{a_2^2}+\frac{1}{a_3^2}
-2\left(\frac{1}{ca_2}+\frac{1}{ca_3}+\frac{1}{a_2a_3}\right)=0,
\quad\quad a_1=\frac{3ca_2a_3}{2(a_2a_3+c (a_2+a_3))}.
\end{equation}
If (\ref{condA}) holds, then the branch points on ${\cal E}_1$ are
$\displaystyle \{ e_1, e_2,e_3\} =
\left\{\frac{2a_1-c}{6a_1c},\frac{2a_1-a_i}{6a_1a_i}, \;\;
i=2,3\right\}$, $\displaystyle \beta=-\frac{1}{2a_1}$ and
$\displaystyle g_2 = -\frac{1}{12a_1^2}$.
\end{proposition}
(\ref{condA}) may be inverted and we get $a_1,a_3$ parametrically
in function of $a_2,c$ or $a_2,c$ in function of $a_1,a_3$.

\begin{corollary}
Let $a_2,c>0$ be given and $a_2\not =c$. Then (\ref{condA}) is
equivalent to \[\displaystyle a_3 = \left( \frac{1}{\sqrt{a_2}}-
\frac{1}{\sqrt{c}} \right)^{-2}, \quad\quad\displaystyle a_1
=\frac{3a_2c}{4(a_2+c-\sqrt{ca_2})}.\]

Let $0<a_1<a_3$ be given. Then (\ref{condA}) is equivalent to
\[\displaystyle \frac{1}{a_2},\quad \frac{1}{c} = \pm
\frac{1}{2\sqrt{a_3}} + \sqrt{\frac{4}{3a_1}-\frac{3}{4a_3}}.\]
\end{corollary}

\paragraph{The second covering and the period mapping}

Let $\Gamma = \{ \mu^2 = -\lambda\prod\limits_{k=1}^4
(\lambda-b_k)\}$, then the second 3:1 covering $\pi_2\, :\,
\Gamma\to {\cal E}_2$ has topological characteristic $(0,3,1,1)$
(see \cite{Enol,Smirnov2}).

In this case the explicit expression of the covering $\Pi_2\; :\;
{\cal G}\mapsto {\cal E}_2$ is known \cite{Smirnov2} and it is
given by the maps \[\displaystyle {\cal Z}=-\frac 14 (4z^3-9 g_2
z-9g_3),\quad\quad \displaystyle {\cal W}=- \frac12 w\left( 4z^2-3
g_2\right)\] and the moduli of ${\cal E}_2$ are $ G_2=\frac
{27}{4}(g_2^3+9g_3^2)$, $ G_3=-\frac {243}{8}g_3(3g_3^2-g_2^3)$.
The finite branch points of ${\cal E}_2$ are $E_1 =-9/2g_2$, $E_2
= 9/4g_3+3/4g_2\sqrt{3g_2}$ and $E_3 = -E_1-E_2$ and satisfy
$E_2<E_1<E_3$.

Using the birational transformation $z=1/\lambda-\sqrt{3g_2}$, we
find the explicit expression of the covering $\pi_2:\Gamma\mapsto
{\cal E}_2$. It is ramified of order 3 at $b_0=0$ (and mapped to
infinity by $\pi_2$), that is $\pi_2^{-1} (E_{\infty})= \{
P_0,P_0,P_0\}$ and
\[
\pi_2^{-1} (E_1) = \{ b_2,b_3,b_4\}, \quad\quad \pi_2^{-1} (E_2) =
\{ b_1, P_{\pm}\} ,\quad\quad \pi_2^{-1} (E_3)= \{ b_{\infty},
Q_{\pm}\},
\]
where $b_{\infty}$ denotes the infinite ramification point of
$\Gamma$, $b_j$s are the finite ramification points of $\Gamma$
(with a slight abuse of notation, we use the same symbol for the
point on the curve and its $\lambda$ coordinate), $P_{\pm}$ and
$Q_{\pm}$ are the real points on $\Gamma$ such that
$\lambda(P_{\pm}) = 2/\sqrt{3g_2}$ and $\lambda(Q_{\pm})
=2/\sqrt{27g_2}$. Finally, it is easy to check that
\[
\lambda(P_{\pm})\in ]b_3,b_4[, \quad\quad \lambda(Q_{\pm})\in
]b_2,b_3[,\] so that
\[
\frac{\displaystyle \oint_{\alpha_1} \omega_2}{\displaystyle
\oint_{\alpha_2} \omega_2} = \frac{\displaystyle \int_{b_1}^{b_2}
\omega_2 }{\displaystyle \int_{b_3}^{b_4} \omega_2
}=\frac{\displaystyle  2\int_{E_2}^{E_1} d{\cal Z}/{\cal
W}}{\displaystyle  4\int_{E_2}^{E_1} d{\cal Z}/{\cal W}}=
\frac{\displaystyle \oint_{\alpha} d{\cal Z}/{\cal
W}}{\displaystyle 2\oint_{\alpha} d{\cal Z}/{\cal W}}
=\frac{1}{2},
\]
and finally, comparing the definition of period mapping
(\ref{permap}) with (\ref{percov}), we conclude that the period
mapping is either $2:1$ or $1:2$.

Also the dual curve $\Gamma^{\prime}$ defined in Theorem
\ref{theodual} is a hyperelliptic tangential cover of degree $d=3$
and the branch points of $\Gamma^{\prime}$ still satisfy
Proposition \ref{e1-e4}, so that the algebraic real closed
geodesics associated to the dual curve have period mapping $2:1$
or $1:2$.

We have thus proven the following

\begin{lemma}
The closed geodesics associated to a real curve $\Gamma$ which is
a 3:1 hyperelliptical tangential cover, either have period mapping
$2:1$ or $1:2$.
\end{lemma}

\medskip

\subsection{Hyperelliptic tangential covers of degree $4$.}

\paragraph{Description of the hyperelliptic tangential covering and reality
problem for doubly--periodic closed geodesics} In this case, we
require $\Gamma$ to be birationally equivalent to ${\cal G}
=\left\{w^2 = -\prod_{i=1}^5 (z - z_i) \right \}$, where
\begin{equation}\label{zeta4}
\begin{array}{l}
\displaystyle z_1 = 6e_j,  \quad\quad z_{2,3}  = -e_k -2e_j \pm
2\sqrt{((7e_j+2e_k)(e_j-e_k)}, \\\displaystyle z_{4,5}  = -e_l
-2e_j \pm 2\sqrt{((7e_j+2e_l)(e_j-e_l)}.
\end{array}
\end{equation}
${\cal G}$ covers the elliptic curve ${\cal E}_1$ with moduli
$g_2, g_3$, ${\cal E}_1 =\{ \; W^2 = 4 Z^3 - g_2 Z -g_3
=4\prod_{s=1}^3 (Z-e_s) \; \}\subset (Z,W) ,$ and the covering is
given by the relations
\[
Z= e_j+
\frac{(z^2-3ze_j-72e_j^2-27e_l2e_k)^2}{4(z-6e_j)(2z-15e_j)^2},
\quad\quad \frac {d Z}{W} =\frac {(2z-3e_j)\, d z}{w}.
\]
The explicit expression of this covering has been found by
Belokolos and Enolski \cite{Enol} (see also \cite{Smirnov2}). The
reality condition for $z_i$, $i=1,\dots,5$ is
$H_j^2=\prod\limits_{k\not =j}^3 (e_j-e_k)>0$, that is $e_j$ is
either $e_1$ or $e_3$ in (\ref{zeta4}).  If $e_j=e_1$ in
(\ref{zeta4}), then $\beta=6e_1$; if $e_j=e_3$ in (\ref{zeta4}),
then $\beta=-e_2 -2e_3 - 2\sqrt{(7e_2+2e_3)(e_3-e_2)}$. In both
cases, we give necessary and sufficient conditions using the
following notation
\[
f_1=1/b_1, \quad f_2 = 1/b_2,\quad f_3 =1/b_3,\quad f_4=1/b_4.
\]

\begin{figure}[htb]
\psfig{file= 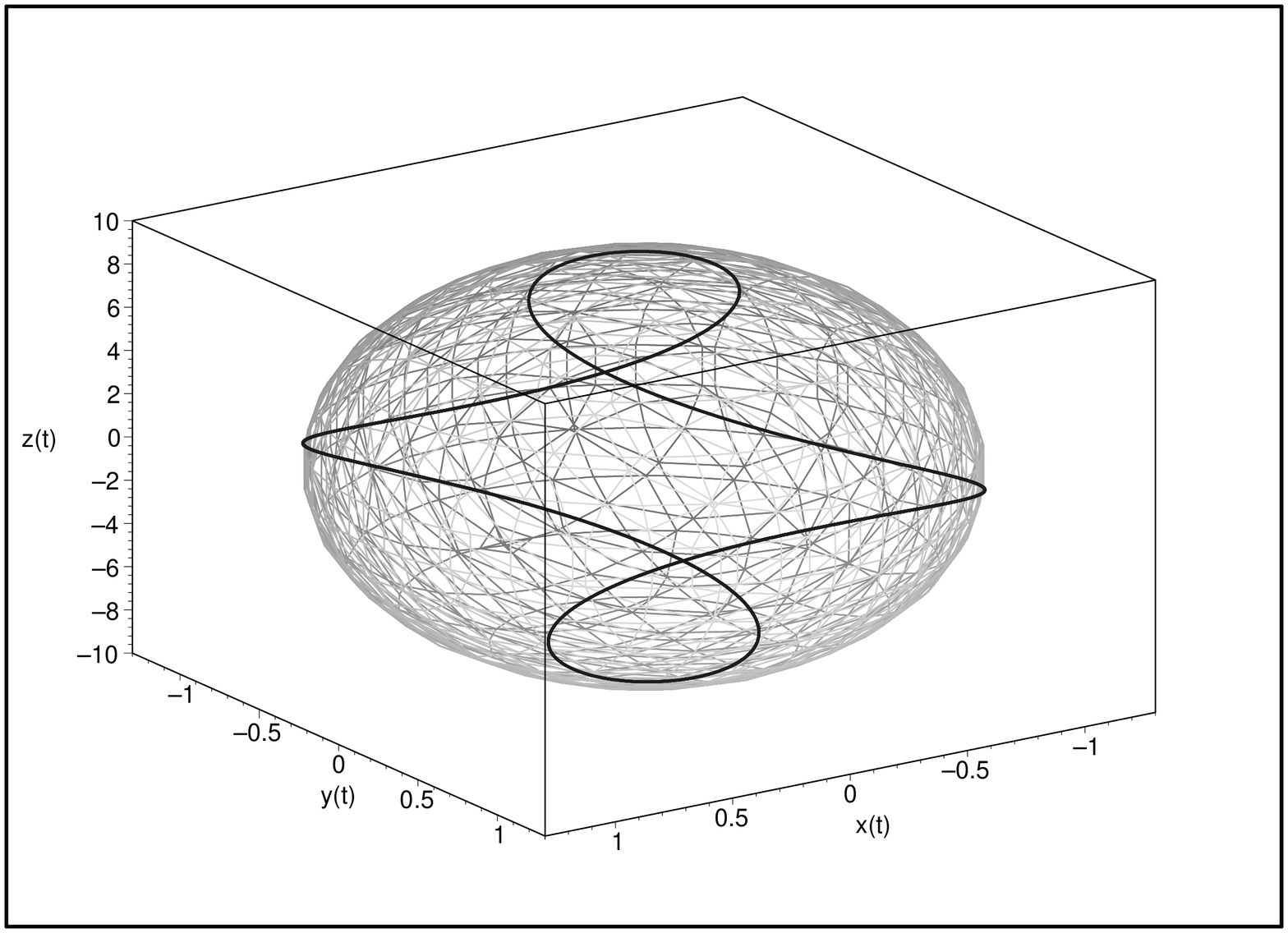,height=6.7cm,width=6.7cm,angle=270}
\vspace{-6.73 truecm}\hspace{6.9 truecm}
\psfig{file=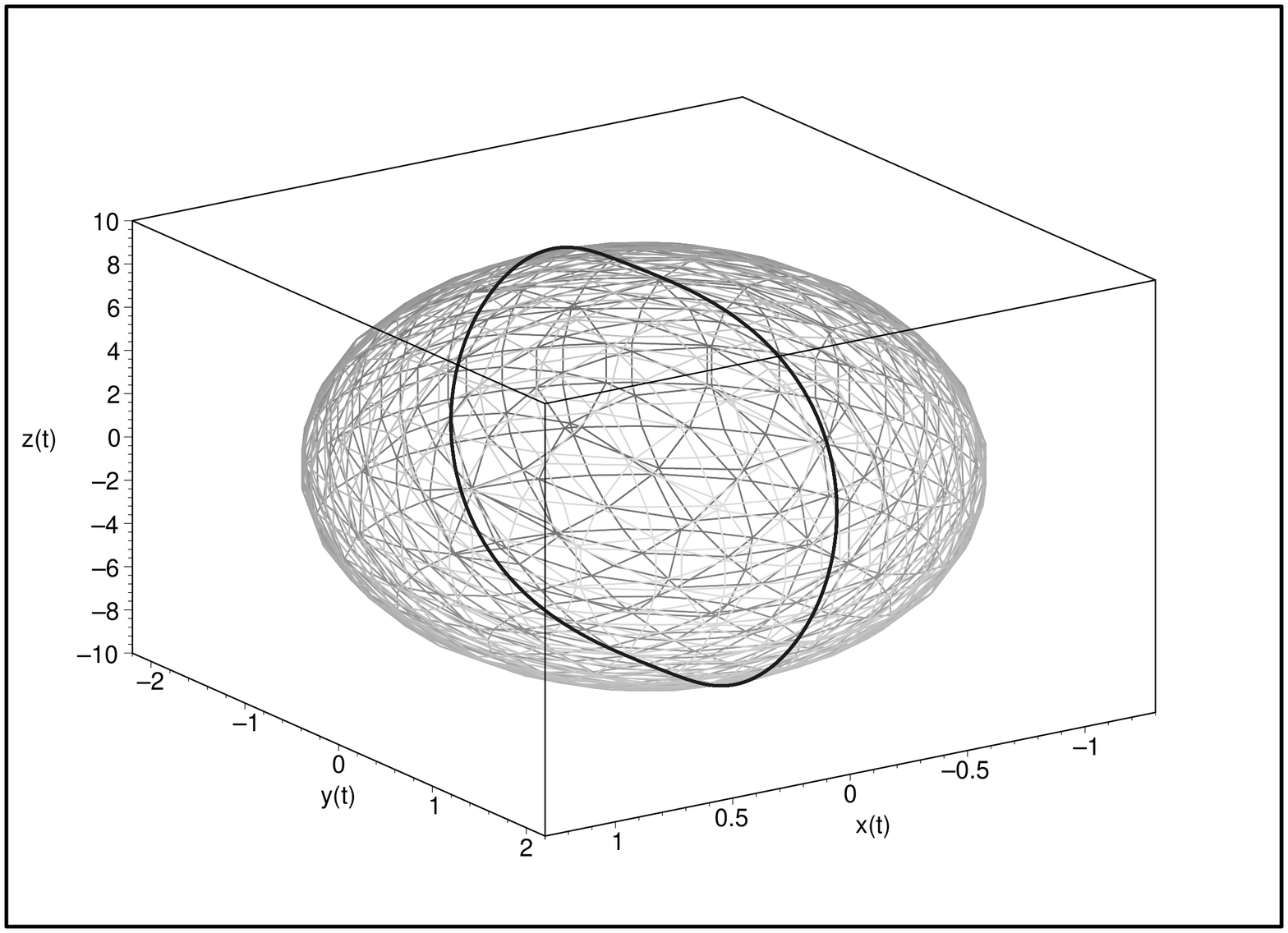,height=6.7cm,width=6.7cm,angle=270}
\centerline{Figure 2.}
\end{figure}

\begin{proposition}\label{betas4}
Let $0<f_4< f_3<f_2 <f_1$ as defined above. Then the geodesic flow
on the ellipsoid $Q$ tangent to the confocal quadric $Q_c$ is
doubly periodic and related, up to birational transformation, to
the 4:1 covering ${\cal G}\to {\cal E}_1$ introduced above, if and
only if either condition (A) or (B) below is fulfilled:
\[
(A) \quad \left\{ \begin{array}{l} \displaystyle 81(f_2^2+f_3^2)
+62f_2^2f_3^2+ 36(f_1+f_4)^2
-108(f_2+f_3)(f_1+f_4)=0,\\
\\
\displaystyle 81(f_1^2+f_4^2) +62f_1^2f_4^2+ 36(f_2+f_3)^2 -
108(f_2+f_3)(f_1+f_4)=0.
\end{array}\right.
\]
\[
(B) \quad\left\{ \begin{array}{l} \displaystyle
100f_4f_3+8f_1(f_4+f_3+8f_1-9f_2)-(9f_4+9f_3-6f_2)^2=0
,\\
\\
\displaystyle (6f_4+6f_3-9f_2)^2+8f_1(9f_4+9f_3-8f_1-f_2)=0
\end{array}\right.
\]
If (A) holds, then $\beta=6e_1$ and the branch points of ${\cal
E}_1$ are \[e_1 = -\frac{1}{30} \left( f_1+f_2+
f_3+f_4\right),\;\; e_2 = \frac{4}{15} \left( f_1+f_4\right)
-\frac{7}{30}\left( f_2+f_3\right),\;\; e_3 =-e_2-e_1\].

\begin{figure}[htb]
\psfig{file= 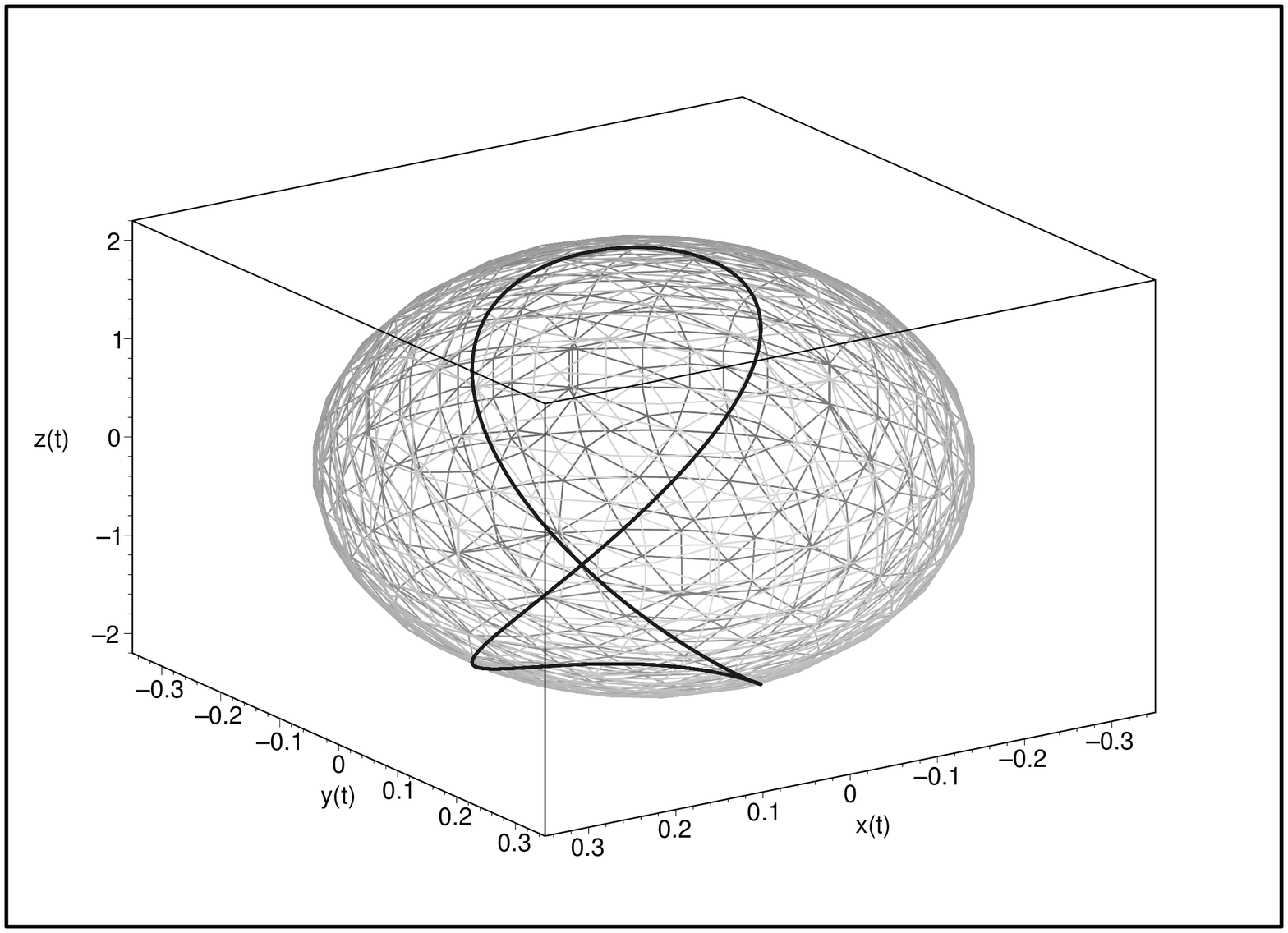,height=6.7cm,width=6.7cm,angle=270}
\vspace{-6.73 truecm}\hspace{6.9 truecm}
\psfig{file=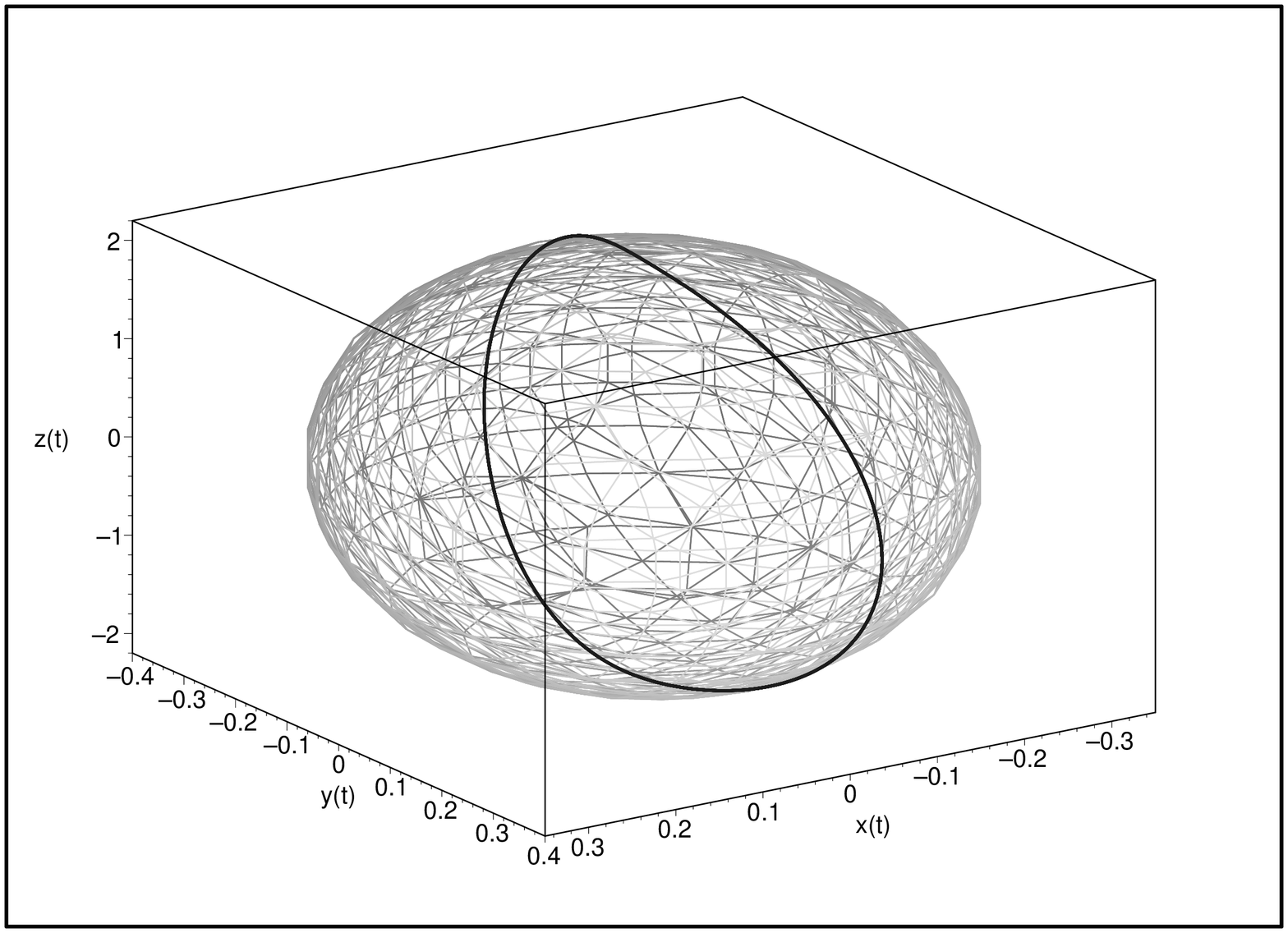,height=6.7cm,width=6.7cm,angle=270}
\centerline{Figure 3.}
\end{figure}

\noindent If (B) holds, then the branch points of ${\cal E}_1$ are
\[
e_3 = \frac{2}{15}f_1-\frac{1}{30} ( f_2+ f_3+f_4), \;\; e_2 =
-\frac{f_1}{15}+\frac{4f_2}{15} -\frac{7}{30}( f_3+f_4),\;\;
e_1=-e_2-e_3\] and $\displaystyle\beta =-\frac{1}{5}
(f_1+f_2+f_3+f_4)$.
\end{proposition}

Conditions (A) and (B) may be expressed in the following
equivalent way

\begin{proposition}\label{betaequal6e1}

$(A^{\prime})$ Let $\sigma_{\pm}>0$ and such that $2/3\sigma_+ <
\sigma_- < \sigma_+$. Let \[\gamma_+ = \frac{9}{25} \left(
\frac{3}{2}\sigma_+ -\sigma_- \right)^2, \quad\quad \gamma_- =
\frac{9}{25} \left( \frac{3}{2}\sigma_- -\sigma_+ \right)^2.\]
Then $0<f_4<f_3<f_2<f_1$ satisfy (A) if and only if $ f_2, f_3$
(resp. $\displaystyle f_1, f_4$) are the roots of \[ x^2-\sigma_+
x +\gamma_+=0, \quad\quad ({\rm resp.} \;\; x^2-\sigma_- x
+\gamma_-=0.\] In this case, the moduli of the elliptic curve are
\[
g_2 = \frac{19}{75} (\sigma_+ +\sigma_-)^2
-\sigma_+\sigma_-,\quad\quad g_3 = \frac{28}{3375}(\sigma_+
+\sigma_-)^3-\frac{\sigma_+\sigma_-}{30} (\sigma_++\sigma_-).\]

$(B^{\prime})$ Let $-\sigma_+<\sigma_-<0$ and define
\[
\gamma_+
=\frac{1}{5}\sqrt{16\sigma_+^2-18\sigma_-\sigma_+-9\sigma_-^2},\quad\quad
\gamma_- = \frac{1}{5}\sqrt{16\sigma_+^2
+2\sigma_+\sigma_--14\sigma_-^2}.\] Then $0<f_4<f_3<f_2<f_1$
satisfy (B) if and only if \[ f_1 = \sigma_+ +\gamma_+, \quad f_2
= 2\gamma_+, \quad f_3 = \frac{\sigma_-}{2} +\gamma_+ +\gamma_-,
\quad f_4 = \frac{\sigma_-}{2} +\gamma_+ -\gamma_-.\] If the above
equation is satisfied, the moduli of the elliptic curve are
\[
g_2=\frac{19}{75}\sigma_-^2-\frac{2}{75}\sigma_-\sigma_++\frac{4}{75}
\sigma_+^2, \quad\quad g_3 = \frac{28}{3375}\sigma_-^3
+\frac{8}{3375}\sigma_+^3 - \frac{37}{1125} \sigma_-^2 \sigma_+
-\frac{2}{1125} \sigma_ -\sigma_+^2.\]
\end{proposition}

\begin{proposition}\label{rem3}  Let $\Gamma = \{ \mu^2 =
-\lambda\prod\limits_{k=1}^4 (\lambda-b_k)\}$ be a real 4:1
hyperelliptic tangential cover verifying Proposition \ref{betas4}
(A). Then the branch points of the dual curve $\Gamma^{\prime}$
introduced in Theorem \ref{theodual} satisfy Proposition
\ref{betas4} (B) and viceversa.
\end{proposition}

\paragraph{The second covering and the period mapping}

Let $\Gamma = \{ \mu^2 = -\lambda\prod\limits_{k=1}^4
(\lambda-b_k)\}$, then the second 4:1 covering $\pi_2\, :\,
\Gamma\to {\cal E}_2$ has topological characteristic $(1,2,2,0)$
and its explicit expression is given in \cite{Enol,Smirnov2}.

Let ${\cal E}_2 = \{ {\cal W}^2 = 4 \prod\limits_{i=1}^3 ({\cal
Z}- E_i)\}$, $E_1<E_2<E_3$, then proceeding as for the case of the
degree 3:1 covering, after some ugly and straightforward
computations, we arrive to the following conclusion. If
Proposition \ref{betas4} (A) holds, $b_0=0$ is a order 3
ramification point mapped to infinity by $\pi_2$, $b_1,b_4 \in
\pi_2^{-1} (E_1)$, $b_2,b_3\in \pi_2^{-1} (E_2)$ and the infinity
point of $\Gamma$ maps to the infinity of ${\cal E}_2$. Finally,
computing the solutions to the equation $(\lambda, \mu)=
\pi_2^{-1} (E_j)$, $j=1,2$ we find real points with $\lambda$
coordinate in $]b_3,b_4[$, and we conclude that
\[
\frac{\displaystyle \oint_{\alpha_1} \omega_2}{\displaystyle
\oint_{\alpha_2} \omega_2} = \frac{\displaystyle 2\int_{b_1}^{b_2}
\omega_2 }{\displaystyle 2\int_{b_3}^{b_4} \omega_2 }= \frac{
\displaystyle 2\int_{E_2}^{E_1} d{\cal Z}/{\cal W}}{\displaystyle
6\int_{E_2}^{E_1} d{\cal Z}/{\cal W} }= \frac{\displaystyle
\oint_{\alpha} d{\cal Z}/{\cal W}}{\displaystyle 3\oint_{\alpha}
d{\cal Z}/{\cal W}}=\frac{1}{3}.
\]
That is, the period mapping is either $3:1$ or $1:3$.

If Proposition \ref{betas4} (B) holds, using Proposition
\ref{rem3} and proceeding as above, we get
\[
\frac{\displaystyle \oint_{\alpha_1} \omega_2}{\displaystyle
\oint_{\alpha_2} \omega_2} = \frac{\displaystyle 2\int_{b_1}^{b_2}
\omega_2 }{\displaystyle 2\int_{b_3}^{b_4} \omega_2 }= \frac{
\displaystyle 2\int_{E_2}^{E_1} d{\cal Z}/{\cal W}}{\displaystyle
4\int_{E_2}^{E_1} d{\cal Z}/{\cal W} }= \frac{\displaystyle
\oint_{\alpha} d{\cal Z}/{\cal W}}{\displaystyle 2\oint_{\alpha}
d{\cal Z}/{\cal W}}=\frac{1}{2}.
\]
and we conclude that the period mapping is either $2:1$ or $1:2$.
We have thus proven

\begin{proposition}\label{permap4}
The closed geodesics associated to a curve $\Gamma$ which is a 4:1
hyperelliptical tangential cover, have period mapping $3:1$ or
$1:3$ in case Proposition \ref{betas4} (A) holds, and have period
mapping $2:1$ or $1:2$ in case Proposition \ref{betas4} (B) holds.
\end{proposition}

\medskip

\paragraph{Figures 2 and 3:}
In figure 2 we present closed geodesics with period mapping $1:3$
($a_1<a_2<c<a_3$) and $3:1$ $(a_1<c<a_2<a_3)$ associated to the
hyperelliptic curve \[\Gamma = \{ \mu^2 =
-\lambda(\lambda-1.453)(\lambda- 1.483)(\lambda-4.434)(\lambda
-84.967)\},\] which is a 4:1 hyperelliptic tangential cover
corresponding to $\sigma_+=2.7$, $\sigma_-=2.1$ so that
Proposition \ref{betas4} (A) is satisfied.

In figure 3 we present closed geodesics with period mapping $1:2$
($a_1<a_2<c<a_3$) and $2:1$ ($a_1<c<a_2<a_3$) associated to the
hyperelliptic curve \[\Gamma = \{ \mu^2 =
-\lambda(\lambda-0.0996)(\lambda- 0.1012)(\lambda-0.150)(\lambda
-4.5510)\},\] which is a 4:1 hyperelliptic tangential cover
corresponding to $\sigma_+=-3$, $\sigma_-=5.1$ so that Proposition
\ref{betas4} (B) is satisfied.

\subsection{Doubly--periodic closed geodesics related to degree 2 coverings with extra
automorphisms}\label{lastexample}

In this section we prove the existence of a family of
doubly--periodic closed geodesics on triaxial ellipsoids
parametrized by $\tau^2\in {\mathbb Q}$ related to the family of
genus two hyperelliptic curves $\Gamma$ which covers 2:1 two
isomorphic elliptic curves ${\cal E}_{1,2}$ (this family of
coverings has also been considered in relation to doubly--periodic
KdV solutions by I. Taimanov \cite{Tai}).

The parameter $\tau$ is the moduli of the elliptic curve ${\cal
E}_1$. Since it is not possible to determine algebraically the
branch points of an elliptic curve in function of the moduli or
viceversa, the condition on $\tau$ is transcendental. However
Theorem \ref{main} implies that for such values of the parameter
$\tau^2$, $\Gamma$ is also a hyperelliptic tangential cover of
another curve ${\cal E}_3$, so that in principle it should be
possible to express such condition also algebraically. Indeed we
have been able to work out an explicit example (Figure 4)
associated to the real intersection of this one parameter family
of degree 2 coverings with extra automorphisms with the
two-parameter family of degree 3 hyperelliptic tangential covers
characterized in subsection 4.1.

\begin{figure}[htb]
\centerline{\psfig{file=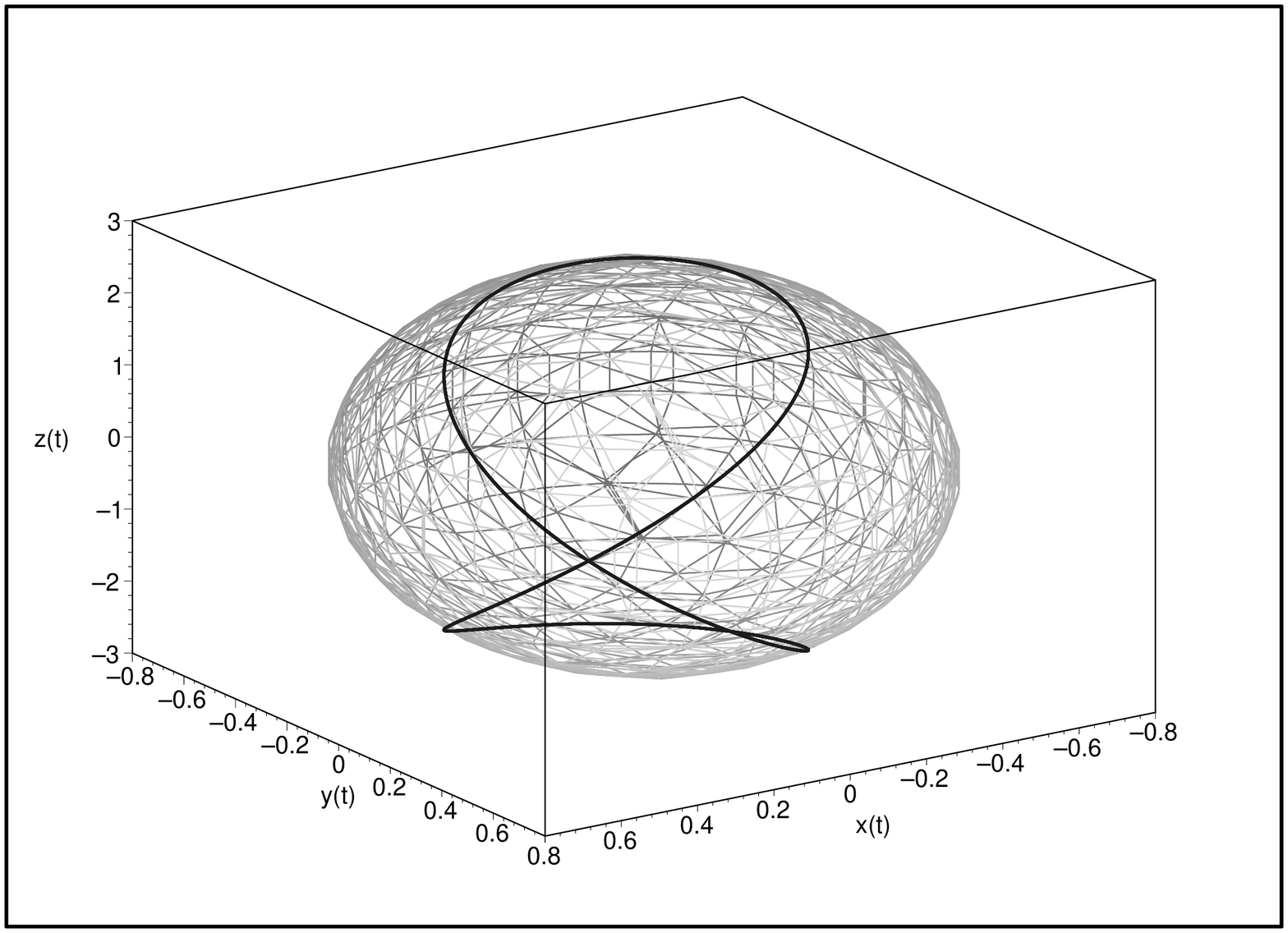,height=7cm,width=7.5cm,angle=270}}
\centerline{Figure 4.}
\end{figure}

\paragraph{Description of the covering} The hyperelliptic curve
\[ \displaystyle {\cal G}_{\alpha} = \left\{\;  w^2 =
z(z^2-\alpha^2) ( z^2 -1/\alpha^2) \;\right\},
\]
covers 2:1 the elliptic curve \[ {\cal E}_1 = \left\{ \; W^2 =
Z(Z-1)(\kappa_{\alpha}^2 Z-1) \; \} \right\},\quad
\kappa_{\alpha}^2 = \frac{(\alpha+1)^2}{2(\alpha^2+1)},\] and the
covering $\Pi_1 \; : \; {\cal G}_{\alpha} \mapsto {\cal E}_1$ is
given by

\[Z = -\frac{2(1+\alpha^2)z}{\alpha(z-\alpha)(z-1/\alpha)}, \;\; W=
\sqrt{-\frac{2(1+\alpha^2)}{\alpha}}
\frac{(z+1)w}{(z-\alpha)^2(z-1/\alpha)^2},\] and, moreover, $
\frac{dZ}{W} = -\sqrt{-\frac{2(1+\alpha^2)}{\alpha}}
\frac{(z-1)dz}{w}$.

\smallskip

\noindent There exists a second 2:1 cover $\Pi_2 \; : \; {\cal
G}_{\alpha} \mapsto {\cal E}_2$, with
\[
{\cal E}_2  = \left\{ \; {\tilde W}^2 = {\tilde A}_{\alpha}
{\tilde Z}({\tilde Z}-1)(\kappa_{\alpha}^2 {\tilde Z}-1) \;
\right\}, \quad\quad {\tilde A}_{\alpha} =
-\frac{2(\alpha+1)^2(\alpha^2+1)}{(\alpha-1)^4},\] \[ {\tilde Z} =
\frac{(z-\alpha)(z-1/\alpha)}{(z-1)^2}, \quad {\tilde W} =
\sqrt{-2\frac{\alpha^2+1}{\alpha}} \frac{y}{(x-1)^3}\] and $
\frac{d {\tilde Z}}{{\tilde W}} =
\frac{(\alpha-1)^2}{\sqrt{2\alpha(\alpha^2+1)}}
\frac{(z+1)dz}{w}$. Clearly ${\cal E}_1$ and ${\cal E}_2$ are
isomorphic since they have the same $j$--invariant (see for
instance \cite{AP}).

Now, let $\alpha>1$. Under the birational transformation,
\[\lambda
= 1/(z+\alpha), \quad\displaystyle\mu =
\frac{y}{\sqrt{1-\alpha^4}(z+\alpha)^3},\] ${\cal G}_{\alpha}$ is
equivalent to
\begin{equation}\label{auto}\displaystyle \Gamma_{\alpha} \; = \;
\big\{ \mu^2 = - \lambda(\lambda -\frac{1}{2\alpha} )(\lambda
-\frac{1}{\alpha} )(\lambda -\frac{\alpha}{\alpha^2-1} )(\lambda
-\frac{\alpha}{\alpha^2+1} )\big\}.\end{equation} Since
$\displaystyle 0<\frac{1}{2\alpha}<
\frac{\alpha}{\alpha^2+1}<\frac{1}{\alpha}<\frac{\alpha}{\alpha^2-1}$,
$\Gamma_{\alpha}$ may be interpreted as the hyperelliptic curve
associated either to the geodesics on the ellipsoid $Q_0$ of
semiaxes $\displaystyle \frac{1}{2\alpha},
\frac{\alpha}{\alpha^2+1},\frac{\alpha}{\alpha^2-1}$ and tangent
to the confocal quadric $Q_c$, $\displaystyle c=\frac{1}{\alpha}$,
or to the geodesics on the ellipsoid $Q_0$ of semiaxes
$\displaystyle \frac{1}{2\alpha},\frac{1}{\alpha}
,\frac{\alpha}{\alpha^2-1}$ and tangent to the confocal quadric
$Q_c$, with $\displaystyle c=\frac{\alpha}{\alpha^2+1}$.

\medskip

The family of hyperelliptic curves $\Gamma_{\alpha}$ is rather
exceptional. Indeed, the birational transformation $\rho =
a_1\lambda/(\lambda-a_1)$ introduced in Lemma \ref{lemmabirat},
just permutes the branch points so that $\Gamma$ coincides with
its dual curve $\Gamma^{\prime}$. Using Theorem \ref{theodual}, we
immediately get

\begin{proposition}\label{propodual}  The real geodesics associated to
$\Gamma_{\alpha}$ are closed if and only if they are
doubly--periodic. In the latter case $\Gamma_{\alpha}$ coincides
with its dual.\end{proposition}

\paragraph{A transcendental condition for doubly--periodic closed geodesics} We
now discuss the existence of such doubly--periodic closed
geodesics for $\Gamma_{\alpha}$. Using the above formulas it is
easy to check that
\[
\frac{dZ}{\sqrt{Z(Z-1)(Z-\kappa_{\alpha}^{-2})}} = 2\rho_{\alpha}
\frac{((\alpha+1)\lambda-1)d\lambda}{\mu},\]\[ \frac{d {\tilde
Z}}{\sqrt{{\tilde Z}({\tilde Z}-1)({\tilde
Z}-\kappa_{\alpha}^{-2})}} = 2i\rho_{\alpha}
\frac{((\alpha-1)\lambda-1)d\lambda}{\mu},\] where $\rho_{\alpha}
= \sqrt{\frac{(1+\alpha^2)}{4\alpha(\alpha-1)(\alpha^2+1)}}$. Now
let $P_1= (\lambda_1,\mu_1),P_2= (\lambda_2,\mu_2)\in \Gamma_a$
and set
\[\displaystyle U_i = 2\int_{\infty}^{\Pi_1 (P_i)}
\frac{dZ}{\sqrt{Z(Z-1)(Z-\kappa_{\alpha}^{-2})}}, \quad
\displaystyle {\tilde U}_i = 2\int_{\infty}^{\Pi_2 (P_i)}
\frac{d{\tilde Z}}{\sqrt{{\tilde Z}({\tilde Z} -1)({\tilde
Z}-\kappa_{\alpha}^{-2})}}, \quad (i=1,2).\] Then, the quadrature
of the geodesics flow
\[
\sum_{i=1}^2 \int_{P_0}^{P_i} \frac{d\lambda}{\mu} = s +const.,
\quad\quad \sum_{i=1}^2 \int_{P_0}^{P_i} \frac{\lambda
d\lambda}{\mu} = const.,
\]
is equivalent to \[ U_1 + U_2 = -\rho_{\alpha} s +c, \quad \quad
{\tilde U_1} + {\tilde U_2} = -\sqrt{-1}(\rho_{\alpha} s + {\tilde
c}),
\]
with $c, {\tilde c}$ constants. Using the addition theorem for
elliptic integrals, the above equations may be inverted and we get
\begin{equation}\label{quada}
{\cal P} ( U_1 | \tau_{\alpha}) + {\cal P} ( U_2 | \tau_{\alpha})
= {\cal P} ( -\rho_{\alpha} s +c | \tau_{\alpha}), \quad {\cal P}
( {\tilde U}_1 | \tau_{\alpha}) + {\cal P} ( {\tilde U}_2 |
\tau_{\alpha}) = {\cal P} ( -\sqrt{-1}(\rho_{\alpha} s +{\tilde
c}) | \tau_{\alpha}),
\end{equation}
where $0<\tau_{\alpha}<1$ is the moduli of ${\cal E}_{1}$. Using
the identity $\displaystyle {\cal P} ( \sqrt{-1}U | \tau_{\alpha})
= {\cal P} \left( U | -\frac{1}{\tau_{\alpha}}\right)$,
(\ref{quada}) is equivalent to
\begin{equation}\label{qqq}
{\cal P} ( U_1 | \tau_{\alpha}) + {\cal P} ( U_2| \tau_{\alpha}) =
{\cal P} ( -\rho s +c | \tau_{\alpha}), \quad {\cal P} ( {\tilde
U}_1 | \tau_{\alpha}) + {\cal P} ( {\tilde U}_2 | \tau_{\alpha}) =
{\cal P} ( -\rho s +{\tilde c} | -\frac{1}{\tau_{\alpha}}).
\end{equation}
Then, the geodesic is doubly periodic if and only if
$\tau_{\alpha}^2 \in {\mathbb Q}$ and, in such a case, the
parameter $s$ may be eliminated from (\ref{qqq}) using the
addition theorem for elliptic functions. In view of theorem
\ref{main}, then $\Gamma_{\alpha}$ also possesses a hyperelliptic
tangential cover of convenient degree $d$ (actually it possesses
an infinite number of coverings following \cite{Pic}). We have
thus proven the following

\begin{theorem}\label{autoth}
Let $\Gamma_{\alpha}$ be the one parameter family of hyperelliptic
curves described above and let $0<\tau_{\alpha}<1$ be the moduli
of ${\cal E}_{1}$. Then, the geodesics associated to
$\Gamma_{\alpha}$ are doubly--periodic if and only if
$\tau_{\alpha}^2 \in {\mathbb Q}$. In the latter case, there exist
an integer $d\ge 3$ and  an elliptic curve ${\cal E}^{(d)}$ such
that $(\Gamma_{\alpha},P_0)$ is also $d:1$ hyperelliptic
tangential cover over ${\cal E}^{(d)}$.
\end{theorem}

From Theorem \ref{autoth} and the argument used to prove
Proposition \ref{propodual}, we immediately conclude the
following.

\begin{corollary}
Suppose that the geodesics associated to $\Gamma_{\alpha}$ are
doubly--periodic. Then the period mapping of the real and
imaginary closed geodesics are either equal or reciprocal to each
other.
\end{corollary}

The above Corollary settles quite restrictive conditions on the
possible hyperelliptic tangential coverings associated to
$\Gamma_{\alpha}$. For instance there cannot exist $d=4$
hyperelliptic tangential coverings associated to
$\Gamma_{\alpha}$, since the curve and its dual would possess
different values of the period mapping for that degree of the
covering (compare Propositions \ref{rem3} and \ref{permap4} for
the $d=4$ hyperelliptic coverings with Proposition
\ref{propodual}). Below we construct explicitly a $d=3$
hyperelliptic tangential covering in the family $\Gamma_{\alpha}$.

\paragraph{Figure 4.} We show an example of doubly--periodic
closed geodesic associated to a covering satisfying Theorem
\ref{autoth}. This example possesses rather exceptional and
intriguing properties. Let $\alpha = \sqrt{2/\sqrt{3}}$ and
$\Gamma_{\alpha}$ as in (\ref{auto}), then $\Gamma_{\alpha}$
covers 2:1 the elliptic curve
\[{\cal E}^{(2)}_1 = \left\{ \; W^2 = Z(Z-1)(\kappa_{\alpha}^2
Z-1), \; \} \right\},\quad\quad {\rm where} \;\; \kappa_{\alpha}^2
= 1/2+2\sqrt{2\sqrt{3}}-\sqrt{6\sqrt{3}}.\] Moreover, the branch
points of $\Gamma_{\alpha}$,
\[a_1=1/(2\alpha),\quad
a_2=\alpha/(\alpha^2+1),\quad a_3=\alpha/(\alpha^2-1),\quad
c=1/\alpha,\] also satisfy (\ref{condA}) in Proposition
\ref{e1-e4}, that is $\Gamma_{\alpha}$ is a degree 3 hyperelliptic
tangential cover over the elliptic curve \[{\cal E}^{(3)}_1 = \{
w^2 = 4z^3-2/9\sqrt{3}z \} .\] We remark that the $j$-invariant of
${\cal E}^{(3)}_1$ takes the exceptional value $1728$, that is the
elliptic curve ${\cal E}^{(3)}_1$ possesses non trivial
automorphisms of order two (see \cite{AP} and references therein).

Finally for the second 3:1 cover we find $G_2=g_2$, $G_3=g_3=0$,
that is ${\cal E}^{(3)}_2 = {\cal E}^{(3)}_1$! As expected the
closed geodesics have period mapping 1:2 (one self--intersection),
if we exchange $c$ and $a_2$ we get period mapping 2:1 and simple
closed geodesics.

\paragraph{Acknowledgements}
I warmly thank A. Treibich for many clarifying discussions
concerning the topological classification of hyperelliptic
tangential coverings. I also warmly thank Yu. Fedorov, E. Previato
and P. van Moerbeke for many helpful discussions during the
preparation of the paper. Finally, I thank F. Calogero, B.
Dubrovin, V. Enols'kii, F. Gesztesy, V. Matveev, A. Smirnov and I.
Taimanov for their interest in this research.

This work has been partially supported by the European Science
Foundation Programme MISGAM (Method of Integrable Systems,
Geometry and Applied Mathematics) the RTN ENIGMA and PRIN2006
''Metodi geometrici nella teoria delle onde non lineari ed
applicazioni''. The figures and the computations in the example
section have been carried out with the help of Maple program.


\begin{thebibliography}{50}


\bibitem{AF} Abenda S., Fedorov, Yu. N. Closed geodesics and billiards on
quadrics related to elliptic KdV solutions. {\it Lett. Math.
Phys.} {\bf 76} (2006),  no. 2-3, 111--134.

\bibitem{AP} Accola R.D.M., Previato E. Covers of tori: genus 2.
{\it Lett. Math. Phys.} {\bf 76} (2006), no. 2-3, 135--161.

\bibitem{Audin} Audin M. Courbes alg\'ebriques et syst\`emes int\'egrables:
g\'eod\'esiques des quadriques. {\it Expo. Math.\/} {\bf 12}, no.3
(1994), 193--226

\bibitem{AMM77} Airault H., McKean H. P., Moser J.
Rational and elliptic solutions of the KdV equation and a related
many-body problem.  {\it Commun. Pure Appl. Math. \/} {\bf 30} ,
(1977), 94--148

\bibitem{Enol} Belokolos E.D., Enol'ski V.Z. Isospectral deformations of elliptic
potentials. {\it Russian Math. Surveys} {\bf 44} (1989),  no. 5,
191--193.

\bibitem{BEBIM} Belokolos E.D., Bobenko A.I., Enol'ski V.Z., Its A.R., and Matveev
V.B.
{\it Algebro-Geometric Approach to Nonlinear Integrable
Equations.\/} Springer Series in Nonlinear Dynamics.
Springer--Verlag 1994.

\bibitem{Cal} Calogero F. Exactly solvable one-dimensional many-body problems.
{\it Lett. Nuovo Cimento} {\bf 13} (1975), 411--415

\bibitem{Chasles} Chasles M. Les lignes g\'eod\'esiques et les
lignes de courbure des surfaces du second degre\'e. {\it Journ. de
Math.} {\bf 11}, 5-20 (1846).

\bibitem{CPP} Colombo E., Pirola G.P. and Previato E. Density of elliptic
solitons. {\it J. reine angew. Math.} {\bf 451} (1994), 161--169.

\bibitem{Dub} Dubrovin B.A. Periodic problems for the Korteweg de
Vries equation in the class of finite--band potentials. {\it
Funct. Anal. Appl.} {\bf 9} (1976), 215-223.

\bibitem{Dub_Nov} Dubrovin B. A., Novikov, S. P. A periodic problem for the
Korteweg-de Vries and Sturm-Liouville equations. Their connection
with algebraic geometry. (Russian) {\it Dokl. Akad. Nauk SSSR},
{\bf  219} (1974), 531---534.

\bibitem{Far} Farkas H.M., Kra I. {\it Riemann surfaces}, Graduate
Texts in Mathematics {\bf 71}, Springer--Verlag (1980).

\bibitem{Fed05} Fedorov, Yu. N. Algebraic closed geodesics on a triaxial
ellipsoid.  {\it Regul. Chaotic Dyn.}  {\bf 10}  (2005),  no. 4,
463--485.

\bibitem{GesWei} Gesztesy F., Weikard R. Elliptic
algebro-geometric solutions of the KdV and AKNS hierarchies - an
analytic approach. {\it Bull. Amer. Math. Soc.} {\bf 35} (1998),
271-317.

\bibitem{Gunn} Gunning R.
{\it it Lectures on Riemann Surfaces. Jacobi varieties.} Princeton
University Press. (1972)

\bibitem{Herm} Hermite C. {\it Oeuvres de Charles Hermite. Vol. III},
Gauthier--Villars, Paris, 1912.

\bibitem{Igusa} Igusa J.-I. Arithmetic variety of moduli for genus
two. {\it Ann. Math.} {\bf 72}, 612--649 (1960).

\bibitem{IM} Its A., Matveev V. The periodic Korteweg de Vries
equation. {\it Funkt. Anal. Pril.} {\bf 9}, (1975), 67--70.

\bibitem{J} Jacobi C. {\it Vorlesungen \"uber Dynamik,
Supplementband.}
Berlin (1884).

\bibitem{Klin} Klingenberg W. {\it Riemannian geometry.} de Gruyter
Studies in Mathematics {\bf 1} (1982).

\bibitem{Knorr} Kn\"orrer H. Geodesics on the ellipsoid.
{\it Invent. Math.\/} {\bf 59} (1980), 119--143

\bibitem{Knorr2} Kn\"orrer H. Geodesics on quadrics and
a mechanical problem of C. Neumann.
{\it J. Reine Angew. Math.}  {\bf 334}  (1982), 69--78.

\bibitem{Krich} Krichever I. Elliptic solutions of the Kadomtsev--Petviashvili
equation and integrable systems on $n$ particles on the line. {\it
Funct. Anal. Appl.} {\bf 14} (1980), 45--54.

\bibitem{L} Lax P. Periodic solutions of the Korteweg de Vries
equation. {\it Commun. Pure Appl. Mathem.} {\bf 28} (1975),
141--188.

\bibitem{MKVM2} Mc Kean H.P., Van Moerberke P. The spectrum of the Hill equation.
{\it Inv. Math.} {\bf 30}, (1975), 217--274.

\bibitem{MKVM} Mc Kean H.P., Van Moerberke P. Hill and Toda
curves. {\it Commun. Pure Appl. Mathem.} {\bf 33}, (1980), 23--42.

\bibitem{Moser} Moser J. Various Aspects of Integrable Hamiltonian Systems. In:
{\it C.I.M.E. Lectures, Bressanone, Italy}, (1978)

\bibitem{Moser2} Moser J. Geometry of quadrics and spectral theory.
{\it The Chern Symposium 1979} (Proc. Internat. Sympos., Berkeley,
Calif., 1979), pp. 147--188, Springer, New York-Berlin, (1980).

\bibitem{Nov} Novikov S.P. The periodic problem for the
Korteweg--de Vries equation. {\it  Funkt. Anal. Pril.} {\bf 8}
(1974), 54--66.

\bibitem{Pic} Picard E. Sur la r\'eduction du nombre del p\'eriodes
des int\'egrales ab\'eliennes et, en particulier, dans le cas de
courbes du second genre. {\it Bull. Soc. Math. Fr.} {\bf 11}
(1883), 25--53.

\bibitem{Poi} Poincar\'e H. Sur la r\'eduction des int\'egrales
ab\'eliennes. {\it Bull. Soc. Math. Fr.} {\bf 12} (1884),
124--143.

\bibitem{Smirnov1} Smirnov A. O.  Elliptic solutions of the KdV equation.
{\it  Math. Z. } {\bf 45}, (1989), 66--73

\bibitem{Smirnov2} Smirnov A. O.  Finite-gap elliptic solutions of the KdV equation.
{\it Acta Appl. Math. } {\bf 36}, (1994),  125--166

\bibitem{Tai} Taimanov I. A. On the two-gap elliptic potentials.
{\it Acta Appl. Math.} {\bf 36}  (1994),  no. 1-2, 119--124.

\bibitem{TV} Treibich A., Verdier J.-L. Tangential covers and sums of
4 triangular numbers. {\it C. R. Acad. Sci. Paris} {\bf 311},
(1990), 51--54

\bibitem{TV2} Treibich A., Verdier J.-L. Solitons elliptiques.
With an appendix by J. Oesterlé. {\it Progr. Math.}, {\bf 88}, The
Grothendieck Festschrift, Vol. III,  437--480, Birkhäuser Boston,
Boston, MA, (1990).

\bibitem{TV3} Treibich A, Verdier J.-L. Rev\^etements exceptionnels et somme de
quatre nombres triangulairs. {\it Duke Math. J.} {\bf 68} (1992),
217--236.

\bibitem{Tr} Treibich A. Hyperelliptic tangential covers, and finite-gap
potentials. (Russian) {\it Uspekhi Mat. Nauk} {\bf 56} (2001), no.
6(342), 89--136; translation in {\it Russian Math. Surveys} {\bf
56} (2001), no. 6, 1107--1151

\bibitem{Ver} Verdier J.-L. New elliptic solitons. {\it Algebraic analysis, Vol. II},
Academic Press, Boston, MA, (1988), 901--910.

\bibitem{Weier} Weierstrass K.
\"Uber die geod\"atischen Linien auf dem
dreiachsigen Ellipsoid. In: {\it Mathematische Werke I}, 257--266

\end{thebibliography}
\end{document}